\documentclass[]{aa520}
\usepackage{graphicx}
\usepackage{rotating}
\input{epsf}
\def\mos{{\footnotesize MOS}\ }
\def\mosone{{\footnotesize MOS1}\ } \def\mostwo{{\footnotesize MOS2}\ }
\def\xmm{{\it XMM-Newton\/}} \def \etal {et al.\ } \def \msol {{\rm\
M}_\odot} \def \kev {\rm keV} \def \betamodel {$\beta$--model} \def
\kT {{\rm k}T} \def \kTX {{\rm k}T_{\rm X}} \def \Omo {\rm \Omega_{\rm
m}}

\def \rc {r_{\rm c}}
\def \dc {\delta_{\rm c}}
\def \rs {r_{\rm s}}
\def \rci {r_{\rm c,in}}
\def \rcut {R_{\rm cut}}
\def \nh {n_{\rm H}}
\def \mp {\rm m_{p}}
\def \tc {\theta_{\rm c}}
\def \fg {f_{\rm gas}}

\begin{document}
      \title{The mass profile of \object{A1413} observed with \xmm:
      implications for the $M$--$T$ relation.  }

      \author{G. W. Pratt  \and M. Arnaud}
      \offprints{G. W. Pratt}

      \institute{CEA/Saclay, Service d'Astrophysique,
                 L'Orme des Merisiers, B\^{a}t. 709,
                 91191 Gif-sur-Yvette Cedex, France
                }
      \date{Received 3 June 2002; accepted 11 July 2002}

\abstract{ We present an \xmm\ observation of A1413, a hot ($\kT =
6.5~\kev$) galaxy cluster at $z=0.143$.  We construct gas and
temperature profiles over the radial range up to $\sim 1700$ kpc.
This radius corresponds to a density contrast $\delta \sim 500$ with
respect to the critical density at the redshift of the cluster, or
equivalently $\sim 0.7 r_{200}$.  The gas distribution is well
described by a $\beta$ model in the outer regions, but is more
concentrated in the inner $\sim 250~{\rm kpc}$.  We introduce a new
parameterisation for the inner regions, which allows a steeper gas
density distribution.  The radial temperature profile does not exhibit
a sharp drop, but rather declines gradually towards the outer regions,
by $\sim 20\%$ between $0.1 r_{200}$ and $0.5 r_{200}$.  The projected
temperature profile is well described by a polytropic model with
$\gamma = 1.07\pm 0.01$.  We find that neither projection nor PSF
effects change substantially the form of the temperature profile.
Assuming hydrostatic equilibrium and spherical symmetry, we use the
observed temperature profile and the new parametric form for the gas
density profile to produce the total mass distribution of the cluster.
The mass profile is remarkably well fitted with the Moore
\etal~(\cite{moore99}) parameterisation, implying a very centrally
peaked matter distribution.  The concentration parameter is in the
range expected from numerical simulations.  There are several
indications that beyond a density contrast $\delta \sim 600$, the gas
may no longer be in hydrostatic equilibrium.  There is an offset with
respect to adiabatic numerical simulations in the virialised part of
the cluster, in the sense that the predicted mass for the cluster
temperature is $\sim 40\%$ too high.  The gas distribution is peaked
in the centre primarily as a result of the cusp in the dark matter
profile.  The X-ray gas to total mass ratio rises with increasing
radius to $\fg \sim 0.2$.  These data strongly support the validity of
the current approach for the modeling of the dark matter collapse, but
confirm that understanding the gas specific physics is essential.
 } \authorrunning{G.W. Pratt
\& M. Arnaud} \titlerunning{An \xmm\ observation of \object{A1413}}
\maketitle
%

\section{Introduction}

The simple first-order formation scenario for galaxy clusters, in
which they grow through the gravitational infall and subsequent
merging of smaller subunits, provides a remarkably good description of
the large-scale properties of these objects.  Within this hierarchical
model, the gas trapped in the potential well of a cluster is heated to
the observed X-ray emitting temperatures by the shocks due to the
formation process; merger features in the gas distribution are then
erased in roughly a sound crossing time ($\sim$ few Gyr), leaving the
gas in hydrostatic equilibrium (HE).

Observation of this gas is a powerful tool for uncovering the physical
characteristics and formation history of a cluster.  Substructure in
X-ray images, combined with optical data, can give clues to the
dynamical state (e.g., Buote~\cite{buote01}).  Direct (temperature
maps) and indirect (hardness ratio maps) methods can give an
indication of where (and if) interactions and mergers are still
occurring (e.g., Markevitch \etal~\cite{mark99}; Neumann
\etal~\cite{dmnetal01}).  In addition, for clusters in reasonably
relaxed state, the assumption of HE and spherical symmetry allow the
derivation of the spatial distribution of both the gas and total
cluster mass by using the information from the X-ray surface
brightness and temperature profiles.  This approach, which is of
fundamental use in cluster studies, has been shown to give masses
which are accurate to about $\pm 20\%$ when applied to simulated
clusters (e.g., Evrard, Metzler \& Navarro~\cite{emn96} [EMN96];
Schindler~\cite{schindler96}).

Numerical simulations based on gravitational collapse are an essential
counterpoint to the observations, being as they are ideal scenarios
with exactly measurable quantities, thus offering a direct comparison
with the real data.  A crucial result from these simulations is the
suggestion that CDM haloes with masses spanning several orders of
magnitude follow a universal density profile independent of halo mass
or cosmology (Navarro, Frenk \& White~\cite{nfw97} [NFW]).  As the
X-ray emitting gas lies in the potential well of the CDM halo, this
suggests that many directly measurable cluster properties should
display self-similarity.  This is observationally testable and indeed,
regularity in the local cluster population has been found in previous
{\it ROSAT\/}, {\it ASCA} and {\it BeppoSAX\/} studies, where the gas
density and temperature profiles of hot, relaxed clusters do appear
similar when scaled to units of the virial radius\footnote{Normally
defined from numerical simulations as the radius of fixed density
contrast $\delta = 200$, or $r_{200}$ (e.g., Evrard, Metzler \&
Navarro~\cite{emn96})} (Markevitch \etal~\cite{mark98}; Neumann \&
Arnaud~\cite{dmnma99}; Vikhlinin, Forman \& Jones~\cite{vikh99}; Irwin
\& Bregman~\cite{irbreg01}; De Grandi \& Molendi~\cite{demol02};
Arnaud, Aghanim \& Neumann~\cite{manadmn02}).  The very existence of
these similarities gives strong support to an underlying universality
in the dark matter distribution, leading to a pleasing convergence
between the observed and simulated properties of galaxy clusters.

However, the temperature profiles in particular have generated much
discussion, as rather different profile shapes have been found for
similar samples observed by the same satellite (e.g., Markevitch
\etal~\cite{mark98}, White~\cite{whi00} [{\it ASCA\/}]; Irwin \&
Bregman~\cite{irbreg01}, De Grandi \& Molendi~\cite{demol02} [{\it
BeppoSAX\/}]).  These studies have been hampered somewhat by both PSF
issues and sensitivity limits.  The former has an inevitable effect on
the spatial resolution and is a possible source of systematic
uncertainty, the derivation of the profiles being potentially
sensitive to the exact correction for the PSF and the detailed
modelling of the non-resolved cooling flow component.  The latter
leads to an inability really to constrain parameters beyond the
supposedly isothermal regime, which is expected, from simulations, to
extend to $\sim 0.5 r_{200}$.  As a direct consequence of this, there
are relatively few galaxy clusters for which sufficiently high quality
data were available for an accurate determination of the total mass
and the corresponding density profile.   Furthermore, any
systematic uncertainty in the shape of the radial temperature
distribution can have a direct effect on the derived mass.  For
example, the temperature profile obtained by Markevitch \etal
(\cite{mark98}) gives mass values that are 1.35 and 0.7 times that
derived assuming isothermality at 1 and 6 core radii respectively. As
a result, the actual form of the density profile is still a largely
untested quantity, at least from an observational point of view.

Clusters can also be used to provide cosmological constraints.  For
any given cosmology and initial density fluctuation, the mass
distribution of virialized objects can be predicted for any given
redshift.  Constraints on cosmological parameters, $\sigma_{8}$ and
$\Omega$, can be found by comparing the predictions with the observed
cluster mass function and its evolution (Perrenod~\cite{perr80}).  For
this, however, a great number of accurate observational masses are
needed.  In the calculation of the observed cluster mass function, the
standard way to overcome the paucity of data is to use average cluster
temperatures, taking advantage of the tight mass-temperature relation
predicted by numerical simulations, where $M \propto T^{3/2}$ (e.g.,
EMN96).  While observations have, for hot clusters at least, recovered
the slope of this relation, observed masses imply a normalisation
consistently lower than found by simulations (e.g., Horner, Mushotzky
\& Scharf~\cite{hms99}; Nevalainen, Markevitch \& Forman~\cite{nmf00};
Finoguenov, Reiprich \& B\"{o}hringer~\cite{frb01}).  However, these
total cluster mass estimates, except in a few cases, required an
extrapolation of the data and the level of the discrepancy is
sensitive to the assumed temperature profile (e.g. see Horner
\etal~\cite{hms99}, Neumann \& Arnaud~\cite{dmnma99}).

\xmm\ and {\it Chandra} offer, for the first time, sufficiently good
spatial and spectral resolution for self-consistent determinations of
global cluster observables such as gas density, temperature and mass
profiles.  We are now observing clusters with unsurpassed clarity.
{\it Chandra\/}, with higher resolution, is the instrument best-suited
for the study of cluster cores.  In the most recent {\it Chandra\/}
study by Allen, Schmidt \& Fabian (\cite{asf01}), mass-temperature
data from 6 clusters are measured up to $r_{ 2500}$, and compared to
the reference simulations of EMN96 and Mathiesen \& Evrard
(\cite{me01}).  Once again, a systematic offset of $\sim 40\%$ is
found between the observed and simulated $M-T$ curves, in the sense
that the predicted temperatures are too low for a given mass.  \xmm,
with its high throughput and large field of view, is the satellite
best-matched for the study of the larger scale structure of these
objects, and for the determination of essential quantities out to a
good fraction of the virial radius.  With this capability it is
possible to test for other effects, such as potential variations of
the
normalisation with radius.

In this paper, we use \xmm\ observations of the relaxed cluster A1413
at $z = 0.143$ to derive the large scale properties to high
resolution, and compare the results to those obtained from both
observations and simulations.  We address several questions which have
been the subject of a large amount of debate in the literature.  In
particular, we compare our temperature profile with previously derived
composite profiles from large samples observed with {\it ASCA\/} and
{\it BeppoSAX\/}, and we compare both the form and normalisation of
our mass profile with that expected from numerical simulations.

Throughout this paper we use $H_0 = 50$ km s$^{-1}$ Mpc$^{-1}$,
and unless otherwise stated,$\Omega_{\rm m} = 1$ and $\Omega_\Lambda = 0$
($q_0 = 0.5$).  In this cosmology, at the cluster redshift of $z =
0.143$, one arcminute corresponds to 196 kpc.

\section{Data analysis and preparation}
   \label{sec:data}
\subsection{Observations}

A1413 was observed in Guaranteed Time for 29.4 ks during {\it
XMM-Newton\/} revolution 182 (2000 December 16).  Calibrated event
files were provided by the \xmm\ SOC. The \mos and pn data were
obtained with the {\footnotesize THIN1} and {\footnotesize MEDIUM}
filters, respectively.  For the pn data set, we extracted single events,
which correspond to {\footnotesize PATTERN} 0, while for the \mos data
sets {\footnotesize PATTERN}s 0-12 were selected.

Dedicated blank-sky data sets, which consist of several high-galactic
latitude pointings with sources removed (Lumb~\cite{lumb}), were used
as background for the whole of this analysis.
These data sets are distributed as calibrated event files which have
already been treated with the {\footnotesize SAS}.  We extract the
background events using the same {\footnotesize PATTERN} selection
criteria as outlined above.  In addition, we transformed the
coordinates of the background file such that they were the same as for
the A1413 data set.  In this way we can ensure that all
source/background products come from exactly the same regions of the
detector, thus minimising detector variations.

\subsection{Vignetting correction}

The method described in Arnaud \etal~(\cite{maetal01}) was used to
correct spectra and surface brightness profiles for vignetting
effects.  Briefly, this method involves weighting each photon with
energy $(E)$, detected at position $(x_j,y_j)$, by the ratio of the
effective area at the detected position $A_{x_j,y_j}(E)$ to the
central effective area $A_{0,0}(E)$.

The background data were treated in the same manner as the source.
Note that the background component induced by cosmic rays (see below)
is not vignetted, but since source and background observations are
treated in the same way, the correction factor is the same and thus
cancels.

\subsection{Background subtraction}

The \xmm\ background, consisting of several components, is both time
and energy-dependent, and so subtraction is a subtle process.
Furthermore it is essential that the subtraction is done correctly,
especially so for extended sources like clusters of galaxies, where
background effects begin to play a role at large off-centre distances
where the surface brightness declines approximately as $r^{-4}$ (e.g.,
Vikhlinin et al.~\cite{vikh99}).

The soft proton background due to solar flares cannot be corrected
for in the normal fashion (e.g., spectral subtraction) as it displays
extreme temporal and flux variability, causing the spectrum to change
rapidly with time. At the moment it can only be removed by excising
all frames above a certain count-rate threshold, the main effect of
which is to considerably reduce the effective exposure time. For
these observations, the $3\sigma$ threshold for each camera was
calculated using the method described in Appendix A, and all frames
not meeting this criterion were rejected. In practice the observation
is very clean. Note however that the pn is considerably more
sensitive to the
flares. The final exposure times were 24163s, 24567s and 10254s for
MOS1, MOS2 and pn cameras, respectively. The blank-sky backgrounds
were cleaned using the same criteria.

The blank-sky background represents effectively the particle induced
background, dominant in the hard X-ray band, which is, both
spatially and temporally, relatively constant.  Nevertheless, this
background is variable at the $\sim 10\%$-level, and so it is
frequently necessary to normalise the background.  We normalise these
observations using the count rate in the [10-12] keV and [12-14] keV
  bands, for \mos and pn respectively, treating each camera
separately.  We varied the normalisation by $\pm 10\%$ to assess any
systematic uncertainties.

However, the blank-sky data set does not necessarily represent the
cosmic X-ray background (CXB), because this is variable across the
sky, especially the soft X--ray component (see Snowden
\etal~\cite{snow97}).  We use the method described in Pratt, Arnaud \&
Aghanim (\cite{pratt}), and Arnaud \etal (\cite{maetal02}) to correct
for the difference of the CXB. An annular region external to the
cluster emission (between 9' and 13' in this case) is used to estimate
the local background.  The normalised spectrum of the same region of
the blank-sky background is then subtracted, giving a difference
spectrum, which can then be scaled according to the size of any
extraction region and subtracted directly from the source spectra.  A
similar procedure is applied to subtract the residual CXB component
for the surface brightness profile (see Arnaud \etal~\cite{maetal02}
for details).

In addition to the above, the pn data were corrected for `out of time
events', which occur when a photon hits the CCD during the read-out
process in the imaging mode.

\section{Morphology}
\label{sec:morph}

\subsection{Image}

\begin{figure}
\begin{centering}
\includegraphics[scale=0.47,angle=90,keepaspectratio]{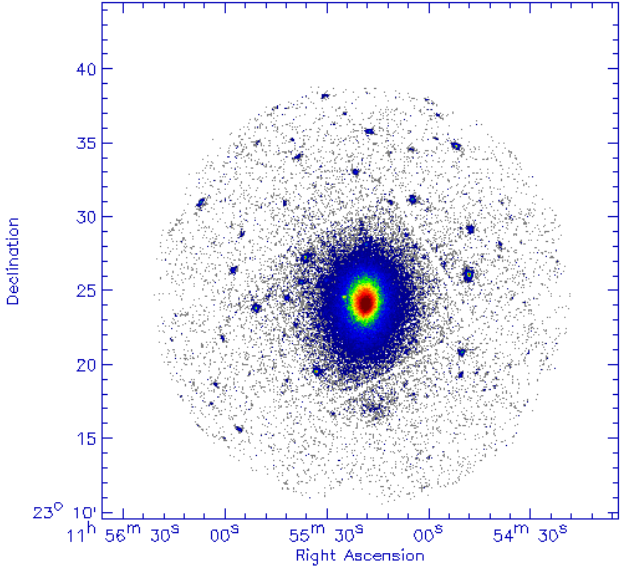}

\caption{{\footnotesize The \mosone+\mostwo counts image of the whole
field of view of the \object{A1413} observation.  Note the large
number of sources and in particular, what appears to the an extended
source to the south of the cluster itself.}}\label{fig:fig1}
\end{centering}
\end{figure}

We show in Figure~\ref{fig:fig1} the {\footnotesize MOS1+MOS2} image
of the cluster, produced simply by adding the data from each camera
without accounting for vignetting. The image is striking: the cluster
displays an unmistakably elliptical shape, with a clear brightness
enhancement directly to the south, and there are a large number of
sources in the field of view.

\subsection{2D $\beta$-model fitting}

Motivated by the apparent excess of counts to the south of the cluster
(see Figure~\ref{fig:fig1}) we fitted the \mosone+\mostwo image with a 2D
$\beta$-model in order to quantify the significance of this feature.
In fitting the image, we followed closely the procedure described in
Neumann \& B\"{o}hringer (\cite{neuboh97}).  Images were extracted in
the [0.3-1.4] keV band from the weighted \mos event files in pixels of
size 3\arcsec.4 and added to make a combined \mos image.  Since in the
case of weighted events Poissonian errors do not apply, errors were
correctly calculated from the weights using $\sigma = \sqrt{\Sigma_{j}
w_j^2}$ (see Arnaud \etal~\cite{maetal01}).  An error image was
generated for each instrument, and these images were added
quadratically.  The fitting procedure described below was tested and
optimised on simulated data before application to the real data.

 The $\chi^2$ test used in the fitting procedure assumes Gaussian
statistics, for which the mean is the most likely value.  The
statistics are actually not Gaussian in the external regions of the
field of view, dominated by the background.  In these regions the
number of photons per pixel is low and follows a Poisson distribution
for which the mean is larger than the most likely value. If the image
is not smoothed before fitting, there is thus a tendency to
underestimate the mean background level, leading to erroneous values
for the fitted cluster parameters.
The combined MOS image was thus smoothed with a
Gaussian of with $\sigma = 10\arcsec$ before fitting.  The error image
was treated according to the error propagation function for Gaussian
filtered images described in Neumann \& B\"{o}hringer
(\cite{neuboh97}).  We fix all error pixels with a value of 0 to have
a value of 1 before fitting, meaning that we can use $\chi^2$ fitting
but are unable to determine confidence parameters on the fit.  The
data were then fitted with a 2D $\beta$-model of the form:

\begin{equation}
S(x,y) = S_0 (1+ F_1 +F_2)^{-3\beta +\frac{1}{2}} + B
\end{equation}
\noindent where
\[
F_1 = \frac{[ \cos{\alpha} (x - x_0) + \sin{\alpha} (y-y_0)]^2}{a_1^2}
\]
\[
F_2 = \frac{[-\sin{\alpha}  (x - x_0) + \cos{\alpha}
(y-y_0)]^2}{a_2^2}
\]

\noindent Here, $x_0,y_0$ is the position of the centre of the
cluster; $x,y$ are the coordinate positions of each pixel; $a_1,a_2$
are the major and minor core radii; $\alpha$ is the position angle;
and the background is included in the model via the parameter $B$.

We fit the image between $1\arcmin.3$ (see below) and $13\arcmin$
from the
cluster center, excluding obvious point sources.  The results of the
2D fit are shown in Table~\ref{tab:2dsbprofs}.  Note that the fitted
parameters are slightly dependent on the outer radius and the $\sigma$
of the Gauss filter, but the results are always in good agreement with
the 1D fit, described below.

\begin{table}
\centering
\center
\caption{{\small 2D $\beta$-model fits, 1'.3 - 13', MOS image}}
\begin{tabular}{l l}
\hline

\multicolumn{1}{l}{ Parameter } & \multicolumn{1}{l}{ } \\
\hline

$\beta$         & 0.72 \\
$\rc$ long & 284.6 kpc \\
$\rc$ short & 201.2 kpc \\
PA              & 2\degr26\arcmin \\
Centre $\alpha$ & 11$^{\rm h}$ 55$^{\rm m}$ 18$^{\rm s}$.9 \\
Centre Dec      & 23\degr 24\arcmin 13\arcsec.8 \\

\hline
\end{tabular}
\label{tab:2dsbprofs}
\end{table}

\begin{figure}
\begin{centering}
\includegraphics[scale=0.50,angle=0,keepaspectratio]{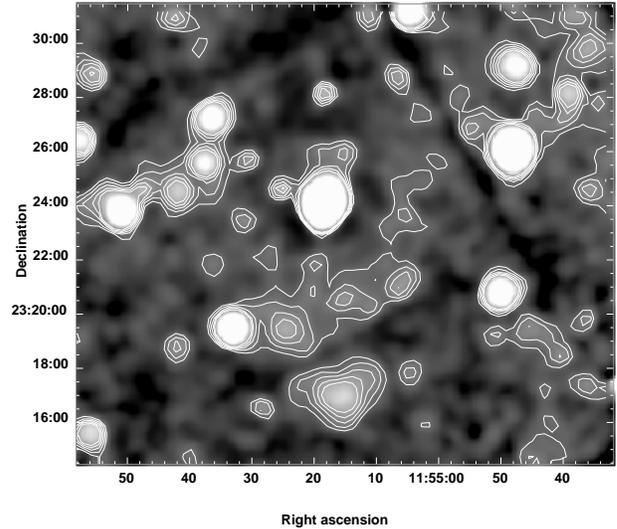}
\caption{{\footnotesize
Residuals after subtraction of the 2D $\beta$-model, smoothed with a
Gaussian with $\sigma = 5/\sqrt{2}$ pixels ($\sim 12\arcsec$).  This
is a zoomed image where the centre of the cluster is the bright
elliptical region at the centre of the image (a possible cooling
flow?).  The
dynamic range is from -10 to +20$\sigma$, where areas of low $\sigma$
are black and areas of high $\sigma$ are white.  Contours
are between +2 and +10$\sigma$, in steps of 2.}}\label{fig:fig2}
\end{centering}
\end{figure}

In order to quantify the significance of the excess of flux to the
south of the cluster, we subtract the 2D $\beta$-model from the data
and calculate the significance of the residuals using the
prescription described in the Appendix of Neumann \&
B\"{o}hringer (\cite{neuboh97}).  The excess is an extended source
detected at
$> 10\sigma$, as shown in  Figure~\ref{fig:fig2}.

We extracted the spectrum from a circular region of radius $\sim
1\arcmin$ centred on the excess. This spectrum unfortunately does not
contain sufficiently strong line emission for a redshift estimate, so
we fitted using a MEKAL model with the same redshift as A1413,
absorbed with the galactic column density toward the cluster
($2.19 \times 10^{20}$ cm$^{-2}$ from Dickey \& Lockman
\cite{dickey}).  We find $\kT = 3.1~\kev$. An overlay of the significance
contours on the DSS plate of the image did not reveal any obvious
sources associated with the excess, and a hardness ratio map did not
reveal any interaction with the main cluster.  Our tentative
conclusion is that the source is either a foreground or background
cluster: deeper optical observations of the region should resolve the
issue.

\section{Gas density profile}
\label{sec:gasden}

\subsection{Surface brightness profile}

For each camera, we generated an azimuthally averaged surface
brightness profile for both source and background observations.
Weighted events from the corresponding event files were binned
into circular annuli centred on the position of the cluster emission
peak. We cut out serendipitous sources in the field of view and
the southern sub-structure.  The background subtraction was performed
as described in Sec.~\ref{sec:data}.  We consider the profiles in
several energy bands.  Due to the contribution of the instrumental Al
K line around 1.5 keV, we ignored the [1.4-2.0] keV band in both
cameras.  To maximise the signal to noise (S/N) ratio, particularly in
the
outer cluster region, we choose to base the following on analysis of
the [0.3-1.4] keV band.

We checked that the vignetting corrected and background subtracted
profiles of the three cameras are consistent: they differ only by a
normalisation factor within the error bars.  We thus coadd the
profiles and bin the resulting profile is the following way.  Starting
from the central annulus, we re-binned the data in adjacent annuli so
that i) at least a $S/N$ ratio of $3\sigma$ is reached and ii) the
width of the bin increases with radius, with $\Delta(\theta) >
0.1\theta$.  Such a logarithmic radial binning insures a $S/N$ ratio
in each bin roughly constant in the outer part of the profile, when
the background can still be neglected.

\begin{figure}
\begin{centering}
\epsfxsize=\columnwidth \epsfbox{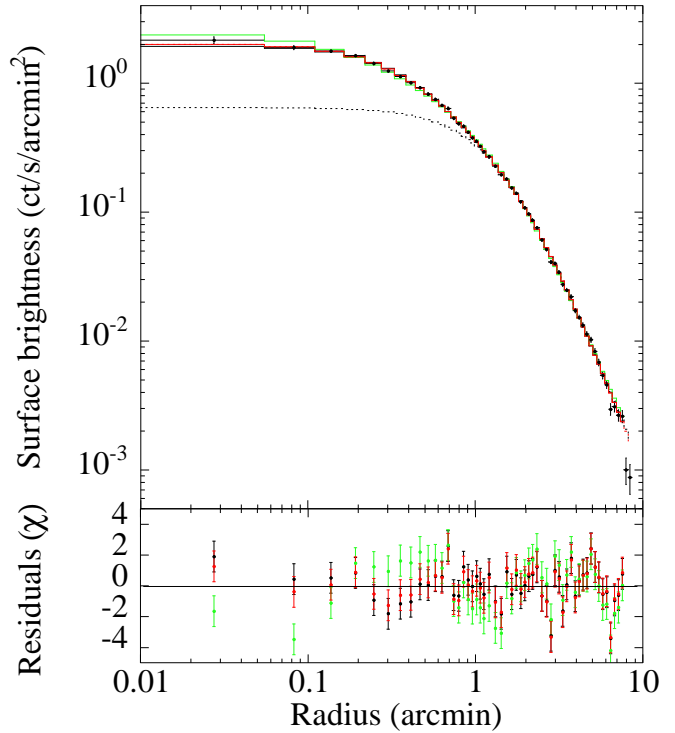} \caption{ Combined
MOS1, MOS2 and pn surface brightness profile of A1413 in the
$[0.3-1.4]~\kev$ energy band.  The profile is background subtracted
and corrected for vignetting effects.  Black (red) [green] lines: best
fit KBB (Eq.\ref{eq:kbb}), BB (Eq.\ref{eq:bb}) and AB (Eq.\ref{eq:ab})
models convolved with the \xmm\ PSF binned as the observed profile.
Dotted line: best fit \betamodel\ fitted to the outer region of the
cluster ($\theta>1.3\arcmin$).  See Sec.\ref{sec:densfit} for model
details and Table~\ref{tab:SXfit} for best fit parameter values.}
\label{fig:FigProffit}
\end{centering}
\end{figure}

The resulting surface brightness profile, $S(\theta)$, is shown in
Fig.~\ref{fig:FigProffit}.  The cluster emission is significantly
detected up to $R_{\rm det}=8.6^\prime$ or $1.7$~Mpc.

\subsection{Density profile modelling}
\label{sec:densfit}

We fitted $S(\theta)$ with various parametric models convolved with
the \xmm\ PSF (Ghizzardi~\cite{ghizzardi}, Griffiths \&
Saxton~\cite{griffiths02b}), binned into the same bins as the observed
profile.

A single \betamodel\ cannot account for the data.  When the entire
radial range is fitted, the reduced $\chi^2$ is $\sim 13$; for the
best fit slope, $\beta=0.60$, and core radius, $\theta_{\rm
c}=0.66\arcmin$.  An excess of emission is readily apparent in the centre
and a lower reduced $\chi^2$ is obtained when excluding the central
region from the fit.  The reduced $\chi^2$ decreases with increasing
cut-out radius until it stabilises for $R_{\rm cut}\sim 1.3'$.  In
that case we obtained $\chi^2=47$ for 31 d.o.f, with
$\beta=0.71\pm0.02$ and $\tc=1.30\arcmin\pm 0.09\arcmin$.  The best
fit model is plotted as a dotted line in Fig.~\ref{fig:FigProffit}.
The $\beta$ value is not surprisingly larger than the value
($\beta=0.62$) derived by Cirimele, Nesci \& Trevese (\cite{cnt97})
from their global fit to the ROSAT profile, but is in excellent
agreement with the value $\beta=0.70\pm0.02$ obtained by Vikhlinin,
Forman \& Jones (\cite{vikh99}) by fitting the outer cluster region
($\theta > 4.9'$).  There is also an excellent agreement between the
1D and 2D $\beta$-values.

We note that the last two points ($7.75\arcmin < \theta < 8.6\arcmin$)
lie significantly below the best fit model (a $3\sigma$ effect for
the last bin).  The cluster flux in the last bin is about 16\% of the
total background and we cannot totally exclude that this discrepancy
is an artifact due to remaining systematic uncertainties in the
background subtraction.  This is further discussed in
Sect.~\ref{sec:dissgasdens}.  These last two points are discarded in
the present analysis.

For the mass analysis which follows (Sect.~~\ref{sec:mass}) it is
convenient to have an analytical description of the gas density radial
profile ($\nh(r)$) at all radii.  We thus tried several alternative
parameterisations, with behaviour at large radii similar to a
\betamodel:

\begin{itemize}
\item {\bf AB model:} A cusped profile similar to the NFW universal
density
profile:

\begin{equation}
\nh(r) = A \left ( \frac{r }{\rc} \right)^{- \alpha} \left [ 1+
\left(\frac{r}{\rc}\right)^{2} \right]^{ \frac{3\beta}{2} +
\frac{\alpha}{2} }
\label{eq:ab}
\end{equation}

where $\alpha$ is the slope at small radii. \\

\item {\bf BB model:} We also introduce a double isothermal $\beta$
model
(BB), assuming that both the inner and outer gas density profile can
be described by a $\beta$-model, but with different parameters.

\begin{equation}
\begin{array}{lllll}
r < \rcut &\nh(r)& = &n_{\rm H,0} &\left[ 1 +
\left(\frac{r}{\rci}\right)^2\right]^{-\frac{3 \beta_{\rm in}}{2}} \\

r > \rcut &\nh(r)& =& N &\left[1 +
\left(\frac{r}{\rc}\right)^2\right]^{-\frac{3 \beta}{2}}\\
   \end{array}
   \label{eq:bb}
\end{equation}
\noindent The boundary between the two regions, $\rcut$, is a free
parameter of the model and we took care that the density distribution
is continious across $\rcut$, as well as its gradient (for continuity
of
the total mass profile, see Eq.~\ref{eq:HE}):
\begin{equation}
N = n_{\rm H,0} \frac{ \left[ 1+ \left( \frac{\rcut}{\rci} \right)^2
\right]^{\frac{-3\beta_{\rm in}}{2} } } { \left[ 1+ \left(
\frac{\rcut}{\rc}\right)^{2} \right]^{-\frac{3\beta}{2} } }
\end{equation}
\noindent and
\begin{equation}
\beta_{\rm in} = \beta~\frac{1  +\left(\frac{\rci}{\rcut}\right)^{2}}
{1  +\left(\frac{\rc}{\rcut}\right)^{2}}
\label{eq:kbb}
\end{equation}

\item {\bf KBB model:} We finally consider a generalisation of the
\betamodel\ for the inner region, allowing a more centrally peaked gas
density profile in the core:
\begin{equation}
\begin{array}{lllll}
r < \rcut &\nh(r)& = &n_{\rm H,0} &\left[ 1 +
\left(\frac{r}{\rci}\right)^{2\xi}\right]^{-\frac{3 \beta_{\rm
in}}{2\xi}} \\

r > \rcut &\nh(r)& =& N &\left[ 1 + \left(\frac{r}{\rc}\right)^2
\right]^{-\frac{3 \beta}{2}}\\
   \end{array}
\end{equation}
\noindent where $\xi<1$ and
\begin{equation}
N = n_{\rm H,0} \frac{ \left[ 1+ \left( \frac{\rcut}{\rci} \right)^{2\xi}
\right]^{\frac{-3\beta_{\rm in}}{2\xi} } } { \left[ 1+ \left(
\frac{\rcut}{\rc}\right)^{2} \right]^{-\frac{3\beta}{2} } }
\end{equation}
\begin{equation}
\beta_{\rm in} = \beta~\frac{1
+\left(\frac{\rci}{\rcut}\right)^{2\xi}}
{1  +\left(\frac{\rc}{\rcut}\right)^{2}}
\label{eq:betai}
\end{equation}
The parameters $\xi$ and $\rci$ are strongly correlated.  An
arbitrary low value of $\xi$ can fit the data if no upper limit is
put on $\rci$.  The lower limit on this parameter given in
Table~\ref{tab:SXfit} is obtained by imposing $\rci<1\arcmin$.
\end{itemize}
\begin{table}
\caption{Results of the surface brightness profile fits.}
\begin{tabular}{l c c c }
\hline
Parameter
& AB model & BB model & KBB model \\
\hline
$n_{\rm H,0}~(10^{-2}{\rm cm}^{-3})$ & - & $2.15$ & $3.07$ \\
$\rc$ &

$1.54\arcmin$&$1.29\arcmin\pm0.10\arcmin$&$1.34\arcmin\pm0.12\arcmin$
\\
$\beta$&
   $0.69$& $0.71\pm 0.02$& $0.71\pm 0.02$ \\
$\rcut$&-&$1.47\arcmin\pm0.13\arcmin$&$1.69\arcmin^{+0.32}_{-0.22}$ \\
$\rci$ & - &- & $0.41\arcmin^{a}_{-0.13}$ \\
$\xi$ & - &- & $0.49^{+0.32}_{-0.16}$$^a$\\
$\alpha$ &0.68&-&-\\
$\chi^{2}$/dof
& $112/51$ & $70.4/48$ &$64.8/47$  \\
$\chi^{2}_{\rm red}$
&$2.20$ & $1.47$ &$1.38$   \\
\hline
\end{tabular}
\label{tab:SXfit}
\smallskip
Notes:   All
errors are at the $90\%$ confidence level.

\noindent
$^{a}$ The maximum value of $\rci$ is fixed to $1\arcmin$.
\end{table}

The corresponding surface brightness profile is computed numerically
by integration of the emission measure along the line of sight:
\begin{equation}
   S_{\rm X}(\theta) \propto
   \int_{r}^{\infty} \frac{\nh^{2}(b)}{ \sqrt{b^{2}-r^{2}}} dr^{2}
   \label{eq:sx}
   \end{equation}
where $r=d_{\rm A}\theta$ and $d_{\rm A}$ is the angular distance.
The emissivity in the considered energy band was estimated using an
absorbed isothermal model at the cluster mean temperature (given
Sect.~\ref{sec:globspec}), taking into account the instrument
response.  In the soft energy band considered, this emissivity is
insensitive to the observed temperature gradient (shown
Sect.~\ref{sec:rtprof}). Note that the profile beyond $\rcut$
obtained for the BB and KBB models is a classical \betamodel.  The
inner surface brightness profile for the BB model can be analytically
computed using incomplete Beta functions.

The best fit models are plotted in Fig.~\ref{fig:FigProffit}, together
with the residuals.  The corresponding best fit parameters with errors
and $\chi^{2}$ values are given in Table~\ref{tab:SXfit}.  In all
cases, the outer slope, $\beta$, is consistently found to be similar
to the slope obtained by fitting only the outer part of the profile.
We found that the AB model does not provide a particularly good
representation of the data: the reduced $\chi^{2}$ is $\chi^{2}_{\rm
red}\sim 2$ and the residual profile below $\rcut$ clearly indicates
that the gas distribution is less peaked than a cusped profile.  In
other words, the gas distribution possesses a core.  The best fit is
obtained with the KBB model, but the reduced $\chi^{2}$,
$\chi^{2}_{\rm red}\sim 1.38$ is still larger than 1.  However, the
residuals are small (at the $3\%$ level on average) and might be due
in part to the observed departure from spherical symmetry.  As the BB
model is a special case of the KBB model ($\xi$ fixed to $\xi=1$), we
can compare both models using a F-test.  The KBB model provides a
better fit than the BB model at the $95\%$ confidence level,
suggesting that the density distribution in the core is indeed more
centrally peaked than for a conventional \betamodel.  This KBB model is
thus adopted for the remainder of the analysis.

\section{Spatially resolved spectroscopy}

\subsection{Global spectrum}
\label{sec:globspec}

For each instrument, a global spectrum was extracted from all events
lying within 5\arcmin.1 from the cluster emission peak.  This radial
range was chosen to maximise the S/N ratio, allowing us to check in
detail the consistency between the three cameras.  Each global
spectrum was fitted with an absorbed MEKAL model with the redshift
fixed at $z = 0.143$.  The normalisation for each instrument was left
as an additional free parameter.  We excluded the energy bins around
the strong fluorescence lines of Ni,Cu \& Zn from the pn spectrum fit.
These lines, present in the background, are not well subtracted by the
procedure described in Sec.~\ref{sec:data} because they do not scale
perfectly with the continuum of the particle-induced background.  In
all fits we used the following response matrices:
m1\_thin1v9q20t5r6\_all\_15.rsp (MOS1), m2\_thin1v9q20t5r6\_all\_15.rsp
(MOS2) and epn\_ef20\_sY9\_medium.rsp (pn).

Fitting the data from all instruments above $0.3~\kev$, with the
absorption fixed at the galactic value of $N_{\rm H} = 2.2 \times
10^{20}$ cm$^{-2}$, we found inconsistent values for the temperature
derived with the \mos and pn cameras: $kT = 6.91_{-0.23}^{+0.23}$ keV
(MOS1), $6.33_{-0.23}^{+0.23}~\kev$ (MOS2) and
$5.76_{-0.19}^{+0.19}~\kev$ keV (pn).  A better agreement between the
cameras, together with a lower $\chi^{2}$ value, are obtained if the
$N_{\rm H}$ value is let free, but then the best fit $N_{\rm H}$
values are significantly lower than the 21 cm value (see
Table~\ref{tab:globfits}).  In other words the data presents an excess
at low energy as compared to an isothermal model absorbed with the
galactic hydrogen column density.  This effect could be due to a true
soft excess component (e.g Durret \etal~\cite{durret}) and/or an
artifact due to remaining calibration uncertainties.  In particular,
it is known that the EPIC-pn and \mos cameras show a relative flux
difference which increases with energy above $4.5~\kev$, resulting in
a \mos spectral slope flatter than the pn (Saxton~\cite{saxton};
Griffiths \etal~\cite{griffiths02a}).

We then performed a systematic study of the effect of imposing various
high and low-energy cutoffs for each instrument.  The $N_{\rm H}$ is
fixed to the 21 cm value.  Having first found that progressive cutting
of the high energy channels had a negligible effect on the derived
temperatures, we then varied the low energy cutoff, for which the
results are shown in Table~\ref{tab:globfits}.

\begin{table}
\centering
\center
\caption{{\small Influence of the low-energy cutoff. Absorption
values in bold are frozen at the galactic value.}}
\begin{tabular}{ l l l l l}
\hline

\multicolumn{1}{l}{ Instrument } & \multicolumn{1}{l}{ Band } &
\multicolumn{1}{l}{ $N_{\rm H}$
} & \multicolumn{1}{l}{ $\kT$ } & \multicolumn{1}{l}{ $\chi^{2}/{\rm
dof}$ } \\
\multicolumn{1}{l}{ } & \multicolumn{1}{l}{ (keV) } &
\multicolumn{1}{l}{ ($\times 10^{20}$ cm$^{-2}$) } &
\multicolumn{1}{l}{ (keV)} \\

\hline

MOS1 & $> 0.3$ & $1.04^{+0.34}_{-0.31}$ & $7.51^{+0.40}_{-0.30}$ &
394.3/395 \\
        & $> 0.3$ & {\bf 2.19}           & $6.91^{+0.23}_{-0.23}$  &
424.2/396 \\
        & $> 0.6$ & {\bf 2.19}           & $7.15^{+0.25}_{-0.25}$  &
386.7/376 \\
        & $> 0.8$ & {\bf 2.19}           & $7.27^{+0.26}_{-0.26}$  &
358.5/363 \\
        & $> 1.0$ & {\bf 2.19}           & $7.20^{+0.30}_{-0.30}$  &
349.5/350 \\

MOS2 & $> 0.3$ & $1.00^{+0.33}_{-0.32}$ & $6.94^{+0.29}_{-0.29}$ &
381.9/401 \\
        & $> 0.3$ & {\bf 2.19}           & $6.33^{+0.23}_{-0.23}$  &
407.7/402 \\
        & $> 0.6$ & {\bf 2.19}           & $6.54^{+0.24}_{-0.24}$  &
374.7/382 \\
        & $> 0.8$ & {\bf 2.19}           & $6.67^{+0.25}_{-0.25}$  &
354.6/369 \\
        & $> 1.0$ & {\bf 2.19}           & $6.67^{+0.29}_{-0.29}$  &
344.1/356 \\

pn   & $> 0.3$ & $0.64^{+0.28}_{-0.28}$ & $6.77^{+0.33}_{-0.33}$ &
836.6/811 \\
        & $ >0.3$ & {\bf 2.19}           & $5.76^{+0.19}_{-0.19}$  &
906.2/812\\
        & $> 0.6$ & {\bf 2.19}           & $6.14^{+0.28}_{-0.30}$  &
818.8/754 \\
        & $> 0.8$ & {\bf 2.19}           & $6.49^{+0.31}_{-0.30}$  &
729.0/714 \\
        & $> 1.0$ & {\bf 2.19}           & $6.85^{+0.36}_{-0.35}$  &
675.2/673 \\
        & $> 1.2$ & {\bf 2.19}           & $7.01^{+0.41}_{-0.40}$  &
636.6/633 \\
        & $> 1.5$ & {\bf 2.19}           & $7.28^{+0.65}_{-0.47}$  &
582.9/572 \\

\hline
\end{tabular}
\label{tab:globfits}
\end{table} %

\begin{figure}[t]
\epsfxsize=\columnwidth \epsfbox{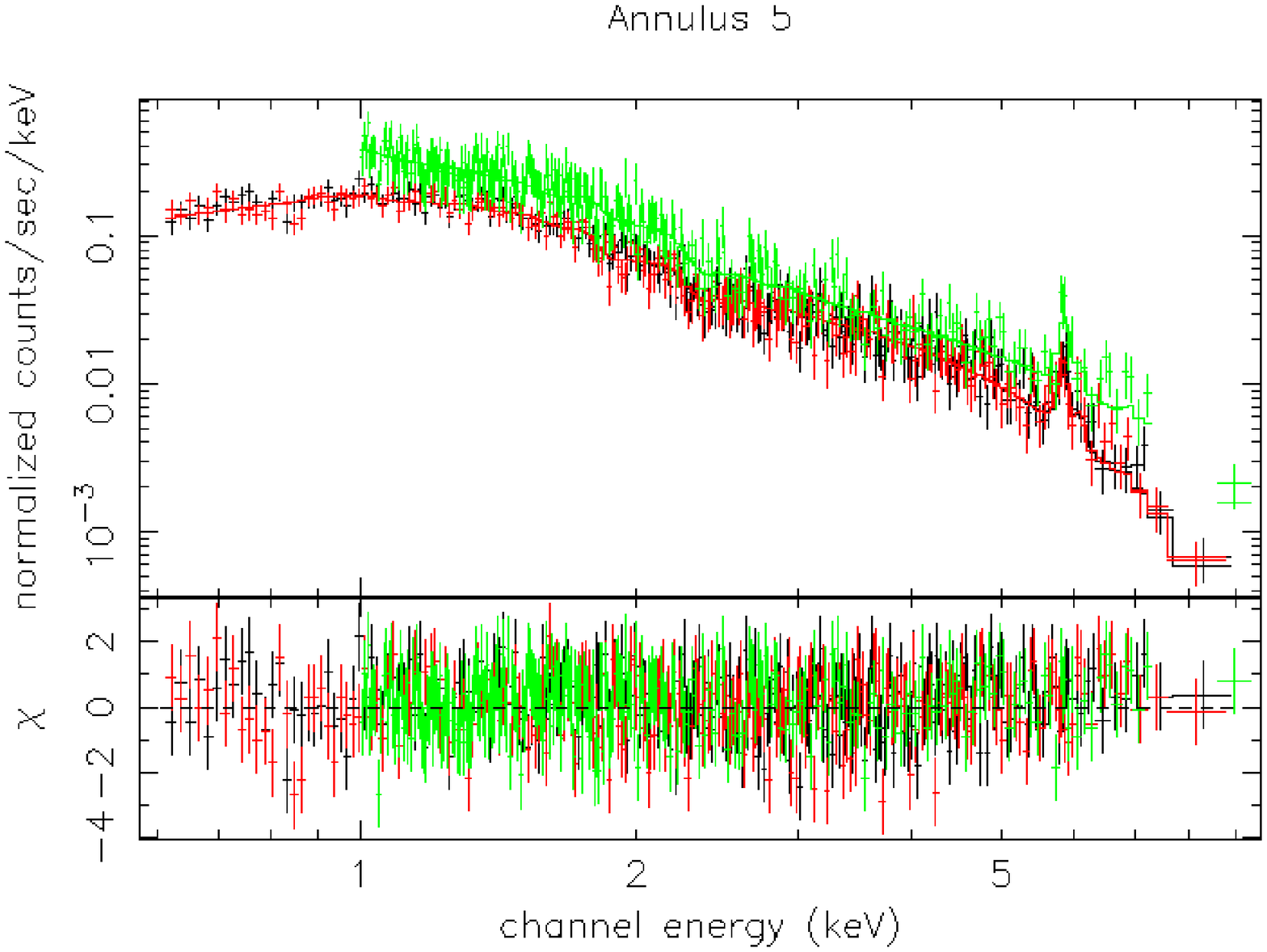} \epsfxsize=\columnwidth
\epsfbox{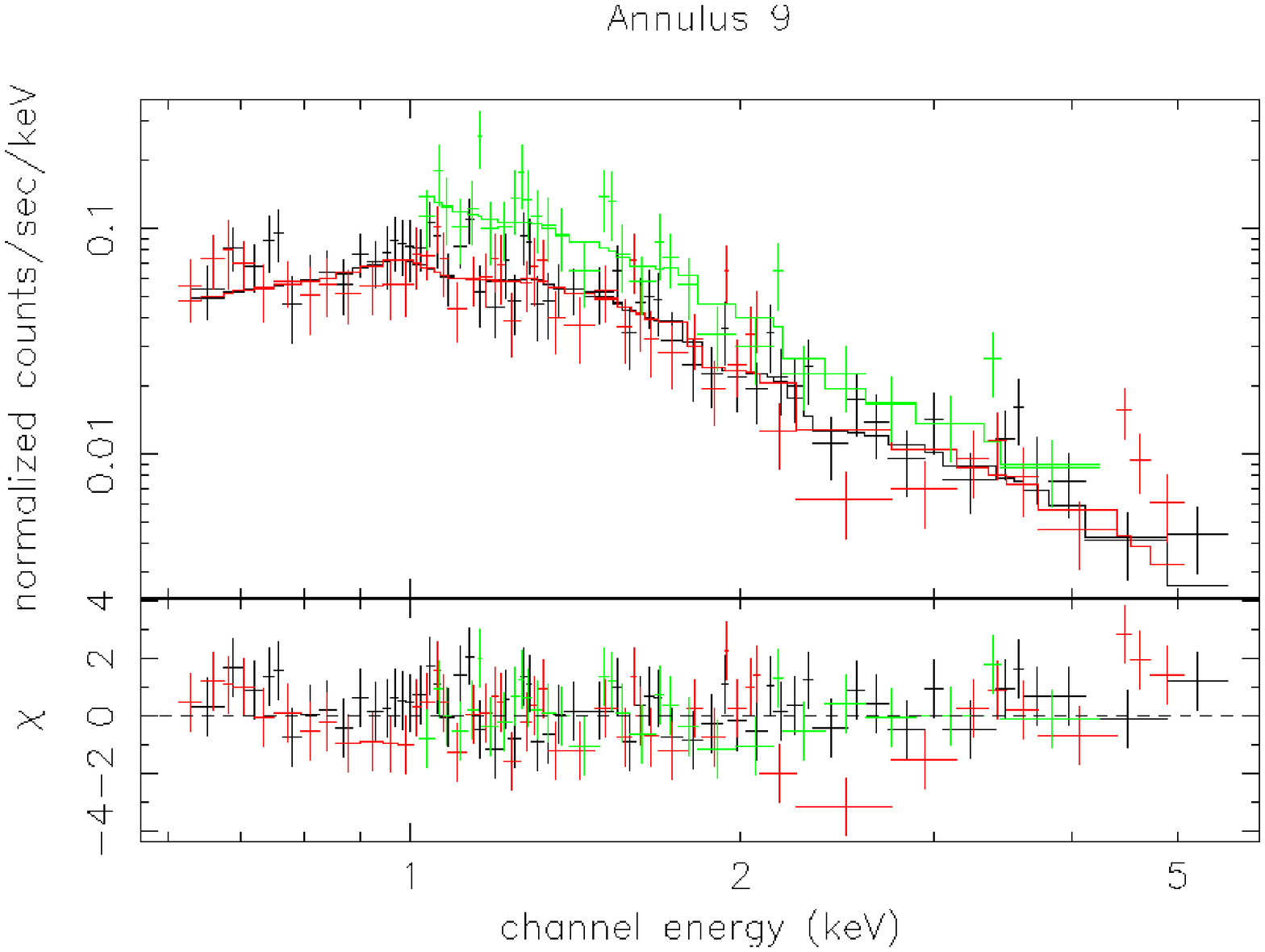} \caption{\xmm\ spectra of the cluster from
annuli 5 ($1.35\arcmin < \theta < 1.89\arcmin$, top panel) and 9
($5.13\arcmin <\theta < 7.16\arcmin$, bottom panel).  Black (red)
[green] points: EPIC/MOS1(2)[pn] data.  The EPIC spectra are
background subtracted and corrected for vignetting as described in
Sec.~\ref{sec:data}.  Solid lines: best fit isothermal model with
parameters given in Table~\ref{tab:ktres}.}\label{fig:annspec}
\end{figure}

This table shows that there is an optimum low-energy cutoff for each
instrument, above which no amount of further cutting of the low-energy
response will significantly affect the temperature.  The temperature
stabilises above a certain cutoff point for each instrument, this
being $\sim 0.6$ keV for the \mos cameras and $\sim 1.0$ keV for the
pn camera.  The adoption of these low-energy cutoffs has the pleasing
effect of bringing the temperatures for each instrument into agreement
both with each other and with previous {\it ASCA\/} analysis.  The
combined \mos+pn global temperature is $\kT = 6.85^{+ 0.15}_{-
0.16}~\kev$ ($90\%$ confidence for one interesting parameter,
$\chi^{2}=1459.6$ for $1436$ dof) in agreement with the results of
Ikebe \etal (\cite{ikebe02}), who find $\kT =
6.56^{+0.65}_{-0.44}~\kev$ and Matsumoto \etal (\cite{matsu01}), who
find $\kT = 6.72 \pm 0.26~\kev$, and marginally consistent with the
result of White (\cite{whi00}), who finds $\kT =
7.32_{-0.24}^{+0.26}~\kev$.

It thus appears that the discrepancies observed by fitting the whole
energy range are mostly due to some residual calibration uncertainties
in the low-energy response of all instruments and/or a true soft
excess.  The scientific analysis of such a possible soft excess is
beyond the scope of this paper.  To minimise these effects, we adopted
the low-energy cutoffs derived above for the spatially-resolved
analysis discussed below.

\subsection{Radial temperature profile}
\label{sec:rtprof}
We produced a radial temperature profile by excluding sources and
extracting spectra in annuli centred on the peak of the X-ray
emission.  All spectra were binned to $3\sigma$ above background level
(except the final annulus, which was binned to $2\sigma$) to allow the
use of Gaussian statistics.   We show the fitted spectra
for annuli 5 and 9 in Figure~\ref{fig:annspec}.

\begin{figure}
\begin{centering}
\epsfxsize=\columnwidth \epsfbox{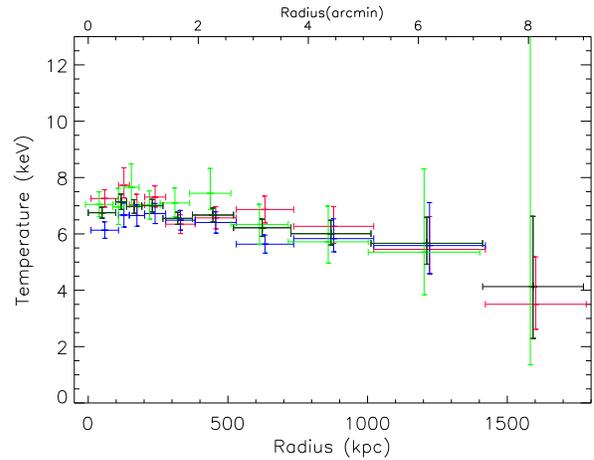} \caption{{\footnotesize
The projected temperature profile of A1413.  The bold black profile is
the total total \mos+pn fit.  For comparison we show the \mos1 (red)
\mos2 (blue) and pn (green) separate fits.  The error bars are
$1\sigma$.}}\label{fig:kTprof}
\end{centering}
\end{figure}

These spectra were fitted using the
absorbed {\sc MEKAL} model described above; we fitted separately the
spectra from each instrument as well as making a simultaneous MOS+pn
fit, as detailed in Table~\ref{tab:ktres}.  All the temperatures are
consistent within the respective errors.  It is also evident from
Figure~\ref{fig:kTprof} that the {\em form\/} of each profile is
similar.  As a further test, we fitted the annular spectra with an
absorbed MEKAL model with the absorption left as a free parameter.
This produced profiles with, again, exactly the same shape, giving us
high confidence in the form of the profile which we have derived.

\subsection{Projection and PSF effects}
\label{sec:deproj}

\subsubsection{PSF correction}

\begin{figure}[t]
\begin{centering}
\epsfxsize=\columnwidth \epsfbox{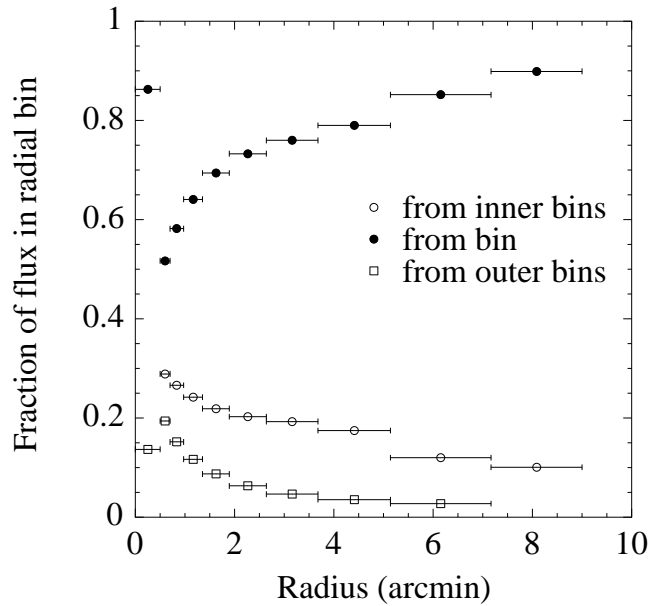} \caption{Redistribution
of the flux due to the \xmm\ PSF: fraction of flux in each radial bin
due to the emission of the bin  (filled circles), as well as all
inner (open circles) and outer bins  (squares).  }
\label{fig:PSFFrac}
\end{centering}
\end{figure}

The PSF of \xmm\ is a potential cause of concern, especially in the
inner regions, where the bin sizes are small.  To assess the effect of
the PSF, we first calculate a redistribution matrix, $F(i,j)$, where
$F(i,j)$ is the fractional flux in annulus $i$ coming from annulus
$j$.  These redistribution factors were derived from our best model of
the gas density profile, converted to emission measure profile and
convolved with the \xmm\ PSF. The fractional contribution in each bin
of the emission coming from the bin, as well as all inner and outer
bins are plotted in Fig.~\ref{fig:PSFFrac}.  The PSF mostly affects
the central regions and, above $2\arcmin$, the contamination from
adjacent bins is less than $25\%$.

We have a total of 30 spectra (10 annuli $\times$ 3 cameras) to be
fitted with a model consisting of 10 MEKAL models
(corresponding to the 10 `true' temperatures) absorbed by a common
(frozen) absorption.  The normalisations of the MEKAL models, for annulus
$i$, are linked by the factors $F(i,j)$, such as to reflect the
contribution of each annulus $j$ due to the PSF. In practice we ignore
any contributions at the less than $1\%$ level.  Each MEKAL model has
6 parameters, which, together with the absorption, makes 7.  If we
fit the 30 spectra simultaneously, the model has $(6 \times 10 +1)
\times 30 = 1830$ parameters.  XSPEC can only handle 1000 parameters,
(even if most of them are frozen), so we have to find a way to reduce
their total number.  One way is to group the spectra, but for
this to work the spectra in each group need a common normalisation.

We typically find differences of $\sim 20\%$ between the
normalisations of MOS and pn annular spectra (\mosone and \mostwo
normalisations are the same to within $\sim 5\%$).  We fit the global
spectrum discussed in Sect.~\ref{sec:globspec} to find the overall
difference in normalisation between \mos and pn.  We then checked that
the annular fit results were the same when the global difference in
normalisation was applied.  This being so, we multiplied the pn annular
spectra by this factor to bring their normalisations into line with
those for the \mos cameras.

We now grouped the \mos and pn spectra of each annulus, giving 10
groups of 3 spectra, leaving us with $(6 \times 10 +1) \times 10 =
610$ parameters, enabling a simultaneous fit.  We froze the
metallicity of each MEKAL model at the best-fit value found for each
projected annulus.  The free parameters in the fit are then the
temperature and normalisation of each MEKAL model.  The resulting
PSF-corrected profile is shown in Figure~\ref{fig:deprpsf} ($1 \sigma$
errors).  The PSF corrected results are entirely consistent with the
projected temperature profile, with systematic differences of about
half the $1\sigma$ errors in the first 3 bins and smaller beyond.
This result is not surprising, since the profile is relatively flat.
 Consideration of the PSF has a much smaller effect on the
temperature profile of A1413 than for the bright cooling flow cluster
A1835 at $z=0.25$ (Markevitch~\cite{mark02}, Majerowicz
\etal~\cite{majerow}).  In contrast to A1835, A1413 displays neither
an extremely steep rise in the central gas density, nor a sharp drop
in the temperature towards the center.  As a result the contamination
of central bins is first smaller: for A1835, more than 1/3 of the
observed brightness at any radius is due to PSF scattering at smaller
radii (Markevitch~\cite{mark02}, Majerowicz \etal~\cite{majerow})
while for A1413 this contamination is already less than $25\%$ at
$2\arcmin$, and decreases beyond (Fig.~\ref{fig:PSFFrac}).  Secondly,
the smaller temperature gradient towards the center means that the
redistribution biases less the temperature determinations.

\subsubsection{Spectral deprojection}

\begin{figure}
\begin{centering}
\includegraphics[scale=0.5,angle=0,keepaspectratio]{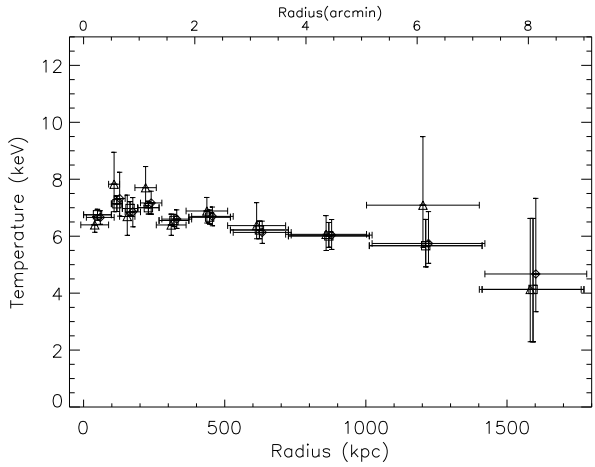}
\caption{{\footnotesize The effect of deprojection and the PSF. The
projected temperature profile of A1413  (squares) compared with the
profile obtained after correction for the PSF  (diamonds) and the
deprojected profile  (triangles).  Errors are $1\sigma$.  }}
\label{fig:deprpsf}
\end{centering}
\end{figure}
Another possible source of error in the derived profile comes from
projection effects.

A deprojected temperature profile was produced by first simultaneously
fitting the \mos and pn spectra of the outer annulus with a MEKAL model
absorbed by a fixed galactic column density.  The spectrum of the next
annulus inward was then fitted with a two-temperature model with the
parameters of one of the models fixed to the best-fitting values
derived for the outer spectrum.  The normalisation of each fixed model
must account for the volume within the outer shell projected along the
line of sight toward the next shell inward.  Furthermore, as the gas
density profile is not flat, the normalisation must also account for
this effect.  We model the gas density profile using the parameters
from the best-fit KBB model described in Sect.~\ref{sec:gasden}.  The
normalisations are then adjusted by the emission weighted volume
factors.  This process was continued inward, adding a MEKAL model for
each annulus, with the parameters of the outer annulus models frozen
to their previously determined best-fit values.  The abundances of the
two outer annuli were frozen to the global value, so for these fits
the free parameter is the temperature.  For all other annuli both the
temperature and abundance were free parameters.  The deprojected
temperature profile is shown compared to the projected profile in
Figure~\ref{fig:deprpsf}.  In practice we find very little difference
between the projected and deprojected results.  The jump in the
temperature of the ninth annulus is somewhat an artifact of the
fitting process.  In this case the software tries to compensate for
the contribution of the low temperature found in the tenth (and
first-fitted) annulus by putting a higher temperature in the
subsequent annular bin.  The error on the tenth temperature is large,
the contribution of the outer emission in the ninth bin depends on the
actual cluster extent and thus the deprojected ninth temperature is
probably more uncertain than found in this simple procedure.  Note,
however, that the projected and deprojected temperatures agree well
within the $1\sigma$ errors.

In summary, we find that neither a consideration of the PSF or
projection effects substantially changes the form of the temperature
profile.  The profiles obtained by taking into account these effects
are consistent with the projected profile, within the $1\sigma$
errors.  For all subsequent analysis, we thus used the projected
profile.

\subsection{Modelling the temperature profile}
\label{sec:tprofmod}

We now consider the scaled temperature profile, $\tau(x)= T(r)/T_{\rm
X}$, where $T_{\rm X}$ is the average temperature and $x$ is the
scaled radius, normalised to $r_{200}$.  $r_{200}$ is estimated from
the average temperature $T_{\rm X}$ and the $r_{200}$--$T$ relation of
EMN96 at the cluster redshift.  $T_{\rm X}$ is estimated by fitting
the global spectrum, extracted from the $[0.5\arcmin-9\arcmin]$
region, i.e outside the possible cooling flow region (see below).  We
found ${\rm k}T_{\rm X} = 6.49\pm0.15~\kev$ ($1\sigma$ error).  Note
that the temperature profile is determined up to $\sim 0.7 r_{200}$ or
$\sim r_{500}$.

We then modelled this projected temperature profile with a polytropic
model:

\begin{equation}
\tau(x) = \tau_0 \left(\frac{\nh(x r_{200})}{n_{\rm
H,0}}\right)^{(\gamma - 1)}
\end{equation}

\noindent where $\nh$ is the gas density profile given by the KBB
model
(Eq.\ref{eq:kbb}), $\gamma$ is the polytropic index and $\tau_0$ is
the normalised temperature at $x=0$.

\begin{figure}[t]
\begin{centering}
\includegraphics[scale=0.5,angle=0,keepaspectratio]{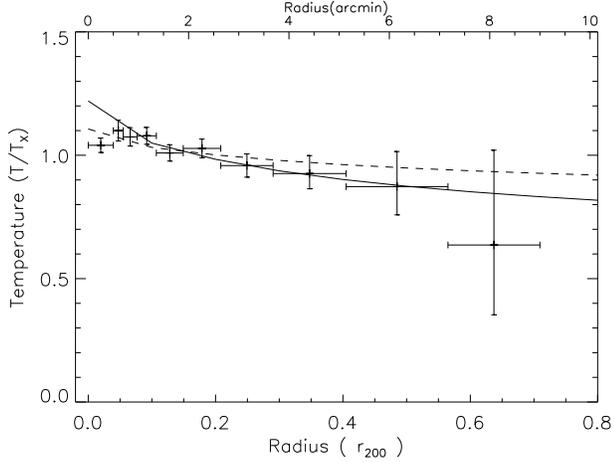}
\caption{{\footnotesize The projected temperature profile of A1413
with the best-fit polytropic model obtained with the central bin
excluded  (full line). The best fit for the entire radial range is
shown as
a dashed line.  }}\label{fig:poly}
\end{centering}
\end{figure}

We fitted the profile with all $\beta$, $\rc$ and $\xi$ parameters
fixed at the values best fitting the surface brightness profile
(Table~\ref{tab:SXfit}), so that the free parameters for the fit are
$\tau_0$ and $\gamma$.  When the whole radial range is fitted, we find
$\tau_0 = 1.11 \pm 0.03$ and $\gamma = 1.03 \pm 0.01$, with $\chi^2 =
7.62/8$.  If we then exclude the inner point, we find a better fit
with $\tau_0 = 1.22 \pm 0.03$ and $\gamma = 1.07 \pm 0.01$, and
$\chi^2 = 4.45/7$.  The polytropic fits to the temperature profile are
shown in Figure~\ref{fig:poly}.

We note that the fit is considerably better if the central point is
excluded from the fit. The resulting polytropic profile rises to
a peak in the centre which is not seen in the annular temperature
determinations, which may lead us to believe that there is a small
cooling flow (CF) at work in the very central regions. This
possibility is discussed further in Sect.~\ref{sec:cengasprop}.

We further note that the derived value for the $\gamma$-parameter is
very close to isothermal and moreover, is not very sensitive to
whether the central bin is included in the fit. There is a
tantalising hint that the temperature profile may drop further in the
very outer regions, but the errors on this last data point are large
enough that it is easily compatible with the derived polytropic model.

\section{Properties of the central gas}
\label{sec:cengasprop}

Our best-fit polytropic model is an excellent description of the
observed temperature profile barring the inner point, which is
significantly lower.  We examined if this temperature drop could be
due to a cooling flow.

The cooling time, the enthalpy of the ICM divided by the energy loss
due to X-rays, is calculated using:
\begin{equation}
t_{\rm cool} =2.9~10^{10}\ {\rm yrs}
\sqrt{\frac{\kTX}{1~\kev}}\ \left(\frac{\nh}{10^{-3}~{\rm
cm}^{-3}}\right)^{-1}
\end{equation}
from Sarazin (\cite{sarazin}).  Using the central density derived from
the KBB model fit (Table \ref{tab:SXfit}), we find $t_{\rm cool} \sim
2.4 \times 10^9$ yr, or about one quarter of the age of the Universe
at the cluster redshift.  This suggests that a CF should exist.
Furthermore, the cooling radius, defined as the radius where the
cooling time is equal to the age of the Universe, is $r_{\rm cool}
\sim 0.6\arcmin$ meaning that any CF should reside in the central bin.
This is consistent with the observed temperature drop.

\begin{table}[t]
\centering
\caption{{\small Multi-temperature and CF fits to the inner annulus.
The $F$-test is computed against the fit for a single temperature
absorbed mekal model}}
\begin{tabular}{ l c c c }
\hline
      Parameters & 1T & 2T & CF \\
\hline
   $\kT_1~(\kev)$ & 6.4 & 6.9 & 7.9 \\
   $\kT_2~(\kev)$ & - & 0.61 &- \\

   $Z/Z_{\odot}$  & 0.33 & 0.35 & 0.35 \\
   $\dot{M}~(\msol~{\rm yr}^{-1})$ & - & - & 58.9 \\
   $\chi^2 / \nu$& 1003.2/854 & 982.2/852 & 978.3/851 \\
   $F_{\rm prob}$& - &$> 99.99\%$ & $> 99.99\%$\\
\hline
\end{tabular}
\label{tab:cffits}
\end{table}
   %
We thus fitted the spectrum of the inner bin with more complicated
models:

\begin{itemize}
\item The sum of two MEKAL models absorbed by a common column density
fixed at the galactic value. When fitting, the second temperature is
limited to be less than or equal to the temperature of the main
component, and the abundances of the two components are tied
together.

\item The sum of a MEKAL and a cooling flow model, again with a fixed
common absorption. Here the abundance of the CF is tied to that of
the thermal spectrum, and the upper temperature for the CF is limited
to be the temperature of the thermal gas.
\end{itemize}

\noindent The results are shown in Table~\ref{tab:cffits}. Both the
two temperature and the MEKAL +CFLOW models are better fits than the
single temperature model at the $> 99.99\%$ level. In addition, the
MEKAL+CFLOW model is a better description of the data than the
two-temperature model at the $93\%$ level.

The secondary temperatures and CF properties are not well
constrained. More sophisticated modelling is needed, preferably
including RGS data, which is beyond the scope of this paper.

\section{Abundance}

\begin{figure}
\begin{centering}
\includegraphics[scale=0.5,angle=0,keepaspectratio]{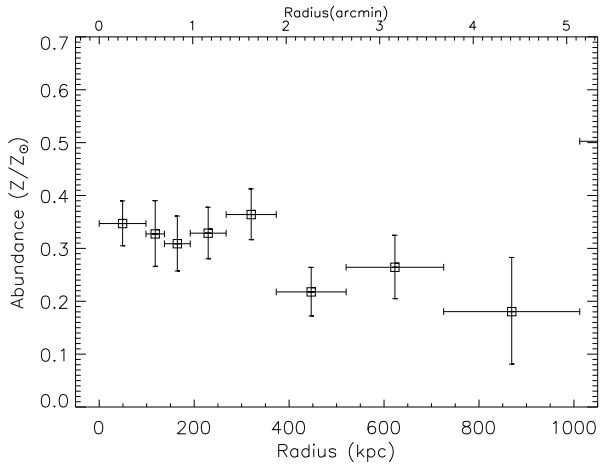}
\caption{{\footnotesize The radial abundance profile of A1413,
derived in the radial range where the spectra have clear detections
of the Fe K line. Errors are $1\sigma$. }}\label{fig:abund}
\end{centering}
\end{figure}

The combined {\footnotesize MOS}+pn spectral fits also allow us to
calculate a radial abundance profile, which can be compared with the
profiles derived for cluster samples observed with {\it BeppoSAX\/}
(Irwin \& Bregman~ \cite{ib01} (IB01); De Grandi \&
Molendi~\cite{dm01} (DM01)).  Fig.~\ref{fig:abund} shows the profile
derived in the conservative radial range where we have information
from the Fe K$\alpha$ line.

The average abundance is $0.27\pm0.03$, more typical of the value
found by DM01 for non-CF clusters
($0.25\pm0.01$) than for CF clusters ($0.34\pm0.01$).  There is a
clear decrease of abundance with radius.  However, the profile
exhibits a two-step behaviour, rather than a steadily declining
profile, as found by DM01 and IB01.  It remains more or less constant
(a mean of 0.34, excluding the central bin) up to $\sim 400$ kpc,
dropping to another plateau (with a mean of 0.22) beyond.  IB01 found
that inside $0.075 r_{200}$ their CF subsample had an average
abundance of $0.46 \pm 0.06$, while the non-CF subsample had an
average of $0.33 \pm 0.04$.  The outer region abundances were $0.30
\pm 0.02$ and $0.24 \pm 0.03$ for the CF and non-CF subsamples,
respectively.  The inner two bins of our observation correspond
roughly to $0.075 r_{200}$.  The mean error-weighted value for this
region is $0.34 \pm 0.08$ (3$\sigma$ errors, for direct comparison
with IB01).  Outside this region, the abundance value is $0.28 \pm 0.09$.
Thus when the errors are taken into account, we cannot distinguish
between the CF and non-CF subsamples of IB01.  However, taken at face
value, these figures appear to suggest that the abundance profile
shape of A1413 displays characteristics intermediate between the CF
and non-CF subsamples.  This may be because A1413 appears to host only
a modest CF. It must also noted that the abundance profiles of A1795
and A2142, both clusters with strong CF signatures (Peres et al.
\cite{p98}), appear to have relatively flat abundance profiles,
as shown in IB01
and DM01. It has been suggested that both of these clusters are
undergoing (or have undergone) mergers (Oegerle \& Hill~\cite{oh94};
Oegerle \etal~\cite{ohf95}), which have presumably not been
sufficiently strong to disrupt the CF, but which have effectively
mixed the metals and thus flattened the radial profile.  A study of
cluster abundance profiles as a function of the strength of the CF
signature and dynamical state would help to better understand the
origin of cluster abundance gradients. 

\section{Total Mass profile}
\label{sec:mass}
\subsection{Calculation of the mass profile}

The mass profile is calculated under the usual assumptions of
hydrostatic equilibrium and spherical symmetry.  The integrated mass
profile can be calculated from the gas density, $n_{\rm g}$, and
temperature profiles:

\begin{equation}
M(r) = - \frac{\kT r}{{\rm G} \mu \mp}   \left[ \frac{d \ln{n_{\rm
g}}}{d \ln{r}} +
\frac{d \ln{T}}{d \ln{r}} \right]
\label{eq:HE}
\end{equation}

\noindent where G and $\mp$ are the gravitational constant and proton
mass and $\mu = 0.597$.

If the gas density profile is
described by the KBB model (Eq.~\ref{eq:kbb}), then the mass profile
is
described by:
\begin{eqnarray}
r < \rcut~~~M(r) & = & - \frac{{\rm k} r^2}{{\rm G} \mu \mp} \left[-
\frac{3 \beta_{\rm in} r^{(2 \xi -1)} T(r)}{r^{2 \xi} + \rci^{2 \xi}}
+ \frac{d T}{d r}\right] \nonumber \\
r > \rcut~~~M(r) & = & - \frac{{\rm k} r^2}{{\rm G} \mu \mp} \left[-
\frac{3 \beta r T(r)}{r^2 + \rc^2} + \frac{d T}{d r}\right]
\end{eqnarray}
\noindent where $\rci$, $\xi$, $\rc$ and $\beta$ are the
parameters of the KBB model, $\beta_{\rm in}$ being linked to them by
Eq.~\ref{eq:betai}.

The mass profile is calculated using the Monte Carlo method of Neumann
\& B\"{o}hringer~(\cite{neubo95}).  The gas density profile parameters
are fixed to their best fit values.  This method calculates random
temperature profiles within the bounds of the observed profile.  We
calculated 10000 such random temperature profiles, using a window size
of 150kpc and a smoothing parameter of $0.1~\kev$.  The final output is
the mean mass profile and the corresponding errors for each data point
in the input temperature profile.  The errors are calculated using the
$90\%$ errors in the temperature profile and the standard deviation of
the mass at any given radius.  The resulting errors are calculated to
correspond to $1\sigma$ errors in the mass profile.

The errors on the mass profile due to the error on the density
gradient, $d \ln{n_{\rm g}}/d \ln{r}$, are then calculated.  As the
parameters $\rci$, $\xi$, $\rc$ and $\beta$ are correlated, this error
cannot be deduced directly from the errors on these parameters.  We
used a method similar to the one described in Elbaz \etal
(\cite{elbaz}).  For each considered radius, the surface brightness
profile is fitted, considering $d \ln{n_{\rm g}}/d \ln{r}$ estimated
at this radius as a free parameter, instead of $\beta$.  The $1\sigma$
error on this parameter is then classically derived from the
$\chi^{2}$ variation, the other parameters (normalisation, $\rci$,
$\rc$ and $\xi$) being optimised.  Finally the errors due to the
density and temperature profiles (derived from the Monte-Carlo method)
are added quadratically.

\begin{figure}
\begin{centering}
\epsfxsize=\columnwidth \epsfbox{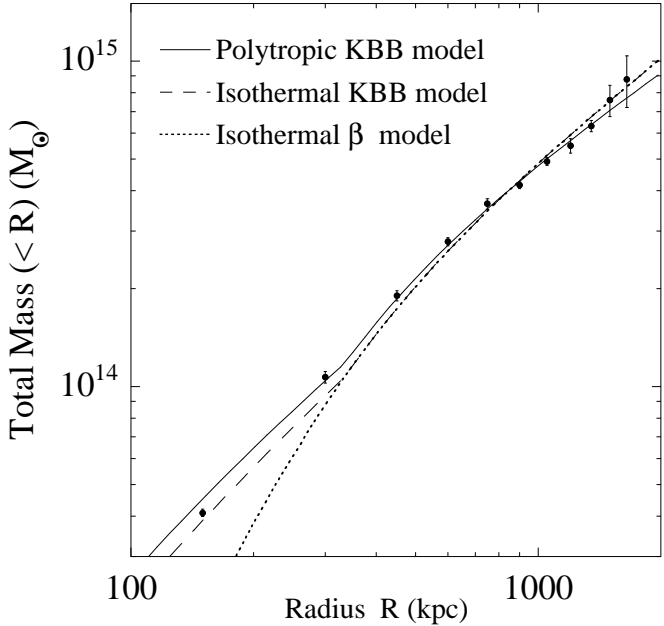} \caption{The mass
profile of A1413 derived from \xmm\ surface brightness and temperature
profiles.  Data points: mass derived from the HE equation, using the
Monte-Carlo method with the best fit KBB model for the gas density
(Eq.~\ref{eq:kbb}) and the observed temperature profile.  Errors bars
are $1\sigma$ and take into account both errors on the temperature and
gas density profiles.  Full line: mass profile derived from the HE
equation and the best fit polytropic model with the KBB model for the
gas density.  Dashed line: same assuming isothermality.  Dotted line:
mass profile derived from the HE equation and an isothermal
\betamodel, fitting the outer gas density profile.
}\label{fig:massmodel}
\end{centering}
\end{figure}

The resulting mass profile, with $1\sigma$ error bars, is plotted in
Fig.~\ref{fig:massmodel}.

\subsection{Factors influencing the mass profile}

The temperature profile of A1413 is well determined out to $\sim 0.7
r_{200}$, and shows a gradual decline which is well described by a
polytrope of index $\gamma =1.07$ beyond the CF region.
Assuming such a polytropic description, the mass profile can be
calculated analytically from the best fit gas density KBB model.  The
polytropic mass profile lies well within the errors of
the Monte Carlo profile, except
for the central point due to the temperature drop observed in the
center (full line in Fig.~\ref{fig:massmodel}).  We also note that the
derived mass at large radii ($r > 1.3$ Mpc) lies at the lower range of
the Monte Carlo mass.  This is due to the drop of temperature in the
last radial bin, the best fit temperature being below the polytropic
value.

In the classic approach, the gas density is described with an
isothermal $\beta$-model, in which the temperature profile is assumed
to be isothermal and the gas density distribution is parameterised by
a \betamodel.  In Fig.~\ref{fig:massmodel} (dotted line) we show the
mass profile obtained using this approach, with the \betamodel\ best
fitting the outer part of the cluster and the average cluster
temperature outside the CF region, $\kTX$.  Not surprisingly, the mass
is greatly underestimated in the centre ($r < \rcut$), where the gas
density profile is more concentrated (higher gradient) than the
extrapolated \betamodel.  If we instead parameterise the gas density
using the best fit KBB model, the mass distribution towards the centre
is recovered (Fig.~\ref{fig:massmodel}, dashed line).  Beyond $\rcut$
the mass profile is slightly steeper than that derived from the true
temperature profile, as expected from the observed $\gamma$ value,
slightly larger than 1.  This comparison shows that the temperature
gradient has a small but systematic effect on the derived mass
profile.

The mass profile is remarkably well constrained: the $1\sigma$ error
is less than $\pm 5\%$ below $1.4$ Mpc and rises to $\sim \pm 18\%$ at
1.8 Mpc.  The temperature logarithmic gradient is much smaller than
the density logarithmic gradient ($7\%$ for $\gamma=1.07$), except in
the very outer part, where the temperature gradient is both larger and
the constraints are poorer.  As a consequence (see Eq.~\ref{eq:HE}),
except in this outer region, the error on the mass profile is
dominated by the error on the gas density gradient (in the range
$0.5\%-3\%$) and on the average temperature ($2.3\%$).  For the same
reason the mass profile is very robust versus possible systematic
errors on the temperature profile.  We have shown in
Sect.~\ref{sec:rtprof} that spectral deprojection or PSF correction do
not have a significant effect on the form of the profile.  One might
also ask what effect the ellipticity of the cluster might have on the
derived radial quantities.  We extracted spectra in elliptical annuli
and compared the projected temperature profile with that produced
using circular annuli.  All temperatures agree within the respective
errors, and so we conclude that the cluster ellipticity is also a
minor source of error.

\subsection{Modelling of the mass profile}
\label{sec:theocomp}

Navarro, Frenk \& White (\cite{nfw97}, NFW) performed high resolution
N-body simulations which showed that the density profiles of dark
matter halos have a universal shape, regardless of halo mass and
values of cosmological parameters.  The NFW profile is given by:
\begin{equation}
\rho(r) = \frac{\rho_{\rm c}(z) \delta_c}{(r/\rs) (1+ r/\rs)^2}
\end{equation}
\noindent where $\rho(r)$ is the mass density and $\rho_{\rm c} (z)$
is the critical density at the observed redshift, which, for a matter
dominated $\Omega = 1, \Lambda = 0$ Universe is:
\begin{equation}
\rho_{\rm c}(z) = \frac{3 H_0^2}{ 8 \pi G} (1 + z)^3.
\end{equation}

The parameters of the model are $\rs$, a scale length and $\dc$, a
characteristic dimensionless density dependent on the formation epoch
of the dark matter halo. $\dc$ can be expressed in term of the
equivalent concentration  parameter, $c$:
\begin{equation}
\dc = \frac{200}{3} \frac{c^3}{[ln (1+c) - c/(1+c)]},
\end{equation}
\noindent The radius corresponding to a density contrast of 200 is
$r_{200}= c\rs$.

\begin{figure}[t]
\begin{centering}
\epsfxsize=\columnwidth \epsfbox{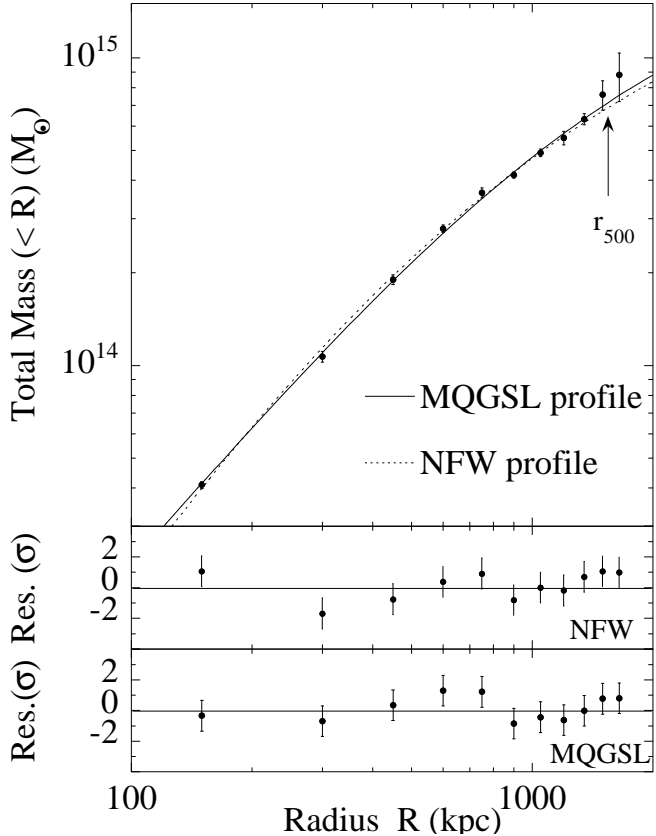} \caption{The mass
profile of A1413 fitted with the NFW profile (dotted line) and Moore
\etal (\cite{moore99}) profile (full line).  Bottom panels: residual
between the data and the model.  The best fit concentration parameters
are $c=5.4$ and $c=2.5$ respectively (see Table~\ref{tab:denfits}).
The radius corresponding to a density contrast of $500$ is indicated
by an arrow.  }\label{fig:massfit}
\end{centering}
\end{figure}

\begin{table}[b]

\centering
\caption{ NFW and Moore et al. (1999) fits to the mass profile
of A1413. Errors are $1 \sigma$}
\begin{tabular}{ l l l l l l }
\hline
Parameter & NFW model & MQGSL model  \\
\hline
   $c$ & $5.4 \pm 0.2$ & $2.6 \pm 0.1$\\
$r_{\rm s}$ (kpc)& $401 \pm 17$& $845 \pm 43$ \\
$r_{200}$ (kpc)& 2169  &  2221\\
$M_{200} (\msol)$ & $8.9 \times 10^{14}$ & $9.5 \times 10^{14}$\\
$\chi^2/{\nu}$ & 8.76/9 &  6.44/9\\
\hline
\end{tabular}
\label{tab:denfits}
\end{table}
   %

The NFW density profile varies from $\rho_{\rm NFW} \propto r^{-1}$
at small radii to $\rho_{\rm NFW} \propto r^{-3}$ at large radii. As
we are fitting the mass profile $M(r)$, we use the integrated mass of
the NFW profile for the fit (e.g Suto, Sasaki \&
Makino~\cite{suto98}):
\begin{eqnarray}
M(r)& = &4\pi \rho_{\rm c}(z) \delta_c \rs^3 m(r/\rs) \\
m(x)& = &ln (1+ x) - \frac{x}{1 + x}
\end{eqnarray}

More recent, higher resolution simulations by Moore \etal
(\cite{moore99},
hereafter MQGSL) suggest a profile described by:
\begin{equation}
\rho(r) = \frac{\rho_{\rm c}(z) \dc}{(r / \rs)^{3/2}
\left[1+\left(r/\rs\right)^{3/2}\right]},
\end{equation}
\noindent where
\begin{equation}
\dc = \frac{100 c^{3} }{\ln {(1 + c^{3/2})}}
\end{equation}
\noindent This is essentially identical to the NFW profile at large
radii but is steeper near the centre ($\rho_{m} \propto r^{-1.5}$).
Again, as we are fitting the mass profile, we use the integrated mass
of the MQGSL profile, given by (Suto \etal~\cite{suto98}):
\begin{eqnarray}
M(r)& =& 4 \pi \rho_{\rm c}(z) \delta_c \rs^3 m(r/\rs)\\
m(x) &=& \frac{2}{3} \ln (1+ x^{3/2})
\end{eqnarray}

The derived parameters from each fit are given in
Table~\ref{tab:denfits} and the best fit models are compared to the
data in Fig.~\ref{fig:massfit}.  We find that the data are extremely
well described by the MQGSL profile across the entire radial range.
The NFW profile can also be used to describe the data, but shows a
small divergence at small radii ($\chi^{2}=4.54$ for the first 3
points).  The radius corresponding to a density contrast of $500$,
computed from the data, is indicated by an arrow.  There is a slight
hint that the measured mass, $M_{500} = 7.7^{+1.2}_{-0.8}\times
10^{14}~\msol$, is higher than the MQGSL and NFW models ($6.8\times
10^{14}~\msol$ and $6.5\times 10^{14}~\msol$, respectively) around
that radius (see Sect.~\ref{sec:mt} for further discussion).


\section{Discussion}

\subsection{The shape of the temperature profile}
\label{sec:tempcomp}

It is instructive to compare the projected \xmm\ temperature profile
of A1413 with the composite profiles found for larger cluster samples.
The most extensive samples come from {\it ASCA\/} and {\it BeppoSAX\/}
data; these are, in order of publication: Markevitch et al.  (1998;
MFSV98), White (2000; W00) Irwin \& Bregman (2000; IB00) and De Grandi
\& Molendi (2002; DM02).  The MFSV98 {\it ASCA\/}-derived profile is
sharply decreasing, such that for a typical $7~\kev$ cluster the
temperature drop is characterised by a polytropic index of 1.2-1.3.
W00 finds that $90\%$ of the cluster profiles in his {\it ASCA\/}
sample are consistent with isothermality at the $3\sigma$-level.  The
IB00 {\it BeppoSAX\/}-derived profile extends only out to $\sim 0.3
r_{200}$ and is flat or even slightly increasing.  In contrast, the
overall DM02 profile, from a larger sample of {\it BeppoSAX\/}
observations,
is characterised by an isothermal core extending to $\sim 0.2
r_{200}$. Their CF subsample exhibits a temperature drop of a
factor of 1.7 between
$\sim 0.2 r_{200}$ and $\sim 0.5 r_{200}$. The non-CF
clusters exhibit a sharper temperature drop in the outer regions.
DM02 suggest that an incorrect treatment of the {\it BeppoSAX\/}
strongback may explain the discrepancy between their result and that
of IB00.

\begin{figure}
\begin{centering}
\includegraphics[scale=0.5,angle=0,keepaspectratio]{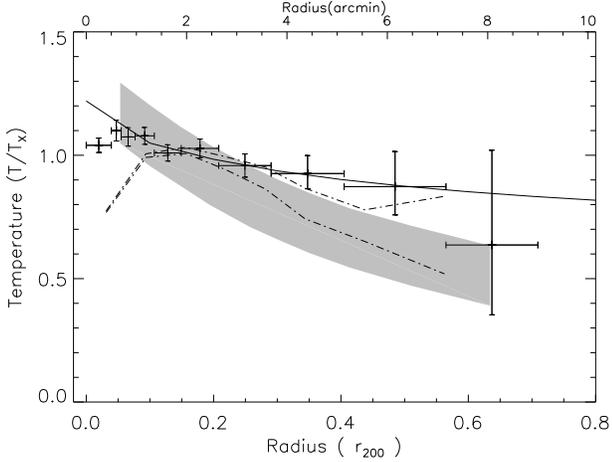}
\caption{{\footnotesize The projected scaled temperature profile of
A1413
compared to the composite CF cluster profile as obtained by De Grandi
\& Molendi~\cite{demol02} (dot-dash lines are joining their data
points plus or minus the $1\sigma$ errors).  The composite profile of
Markevitch \etal~\cite{mark98} is shown as the shaded region
(enclosing the scatter in their best fit profiles).  The solid line is
our best fit polytropic model (excluding the CF region).
}}\label{fig:tempcomp}
\end{centering}
\end{figure}
We have derived the projected temperature profile of A1413 out to
$\sim 0.7 r_{200}$ or $\sim r_{500}$, in much finer detail than is
possible with either {\it ASCA\/} or {\it BeppoSAX\/}.  All
indications are that A1413 is a relaxed cluster.

Our data are compared to the DM02 and MFSV98 composite profiles in
Fig.~\ref{fig:tempcomp}.  Although each individual data point is
(marginally) consistent with the typical region defined by these
composite profiles, there is an obvious systematic difference in
shape.  The A1413 profile does not decline sharply like the composite
profile of MFSV98, or the profile of DM02 beyond $\sim 0.2 r_{200}$.
In Sect.~\ref{sec:tprofmod}, we show that the polytropic model gives
an acceptable fit to the data.  Excluding the central bin, the
$\gamma$ value (1.07) implies an almost isothermal temperature
profile, and is not compatible with that found by MFSV98.  It is very
similar to that found by DM02 for CF clusters, but DM02 reject the
polytropic model on the grounds of poorness of fit, which is not
surprising given the decline of a factor of 1.7 in the temperature of
their composite profile between $0.2$ and $0.5 r_{200}$.  We do not
see a similar decline, and so a polytropic model is a good fit to
these data.  On the other hand, their best fit broken line model is a
poor fit to our data: we find a $\chi^2 = 22.1/6$ for their CF
best-fit, and the fit is worse for their non-CF relation ($\chi^2 =
27.7/7$).

It is obvious that, given the extra radial range afforded by these
\xmm\ data, a flat or increasing profile, such as that of W00 or IB00
extrapolated to high radii, does not describe the A1413 data either.
We emphasize again that the temperature gradient is modest: the
temperature decreases by $\sim 15(20)\%$ between $0.1~r_{200}$ and
$0.3(0.5)~r_{200}$.   Fitting the temperature profile up to $0.3
r_{200}$ (i.e., excluding the last three temperature bins, and the
inner bin) with a polytrope allows us to compare our profile directly
with that of IB00.  We find $\tau_0 = 1.19 \pm 0.03$, $\gamma = 1.06
\pm 0.01$, consistent with the value derived for the full radial
range.  This gradient is in agreement with the level of isothermality
found by W00 and IB00 in that radial range taking into account their
errors, as well as that found by Allen \etal (\cite{asf01}) from
Chandra data below $r_{2500}\sim0.3~r_{200}$.   Our observation is
also consistent with other \xmm\ observations of nearby clusters, e.g.
the slightly decreasing \xmm\ temperature profile of Coma ($10~\%$ at
$0.2~r_{200}$, Arnaud \etal~\cite{arnaud01}) and the temperature
profile of A1795, measured up to $0.4~r_{200}$ and found to be flat
within $\pm 10\%$ beyond the CF region ($0.1~r_{200}$, Arnaud
\etal~\cite{maetal01}).

\subsection{Shape of the total mass profile}

In Sec.~\ref{sec:theocomp} we showed that the NFW form can describe
the total mass profile of A1413.  However a slightly better agreement
in the center is obtained with a MQGSL profile, derived from higher
resolution simulations.

With {\it Chandra\/}, it is possible to examine the central regions in
great detail, at the expense of information at large radii.  At
present, it is unclear whether the NFW or MQGSL profiles provide the
better description of the mass profiles derived from {\it Chandra\/}
observations.  Allen, Schmidt \& Fabian (\cite{asf01b}) investigate
several forms for the mass profile of RXJ1347.5-1145, finding that
both the NFW and MQGSL provide an acceptable fit, although the NFW
profile is favoured in terms of $\chi^2$.  Perhaps the highest
resolution examination of a cluster mass profile is that of Hydra A by
David \etal (\cite{david01}), who find $\rho \propto r^{-1.3}$ between
30 and 200 kpc, which is intermediate between the NFW and MQGSL forms.
The addition of a mass point from H$_{\alpha}$ observations leads them
to favour the NFW profile, although the result is still consistent
with the MQGSL result.

One sticking point is the value of the concentration parameter from
the NFW fit by David \etal (\cite{david01}).  They find $c = 12$,
which is 3 times that expected for a cluster of the mass of Hydra A.
Interestingly, a similar value of $c$ was found by Arabadjis, Bautz \&
Garmire (\cite{abg02}) from a {\it Chandra\/} study of EMSS 1358+6245.
On the other hand, the c parameters of Allen \etal (\cite{asf01}) are
better in agreement with the theoretical predictions.  At large radii
the NFW and MQGSL profiles coincide and we can compare the c value
derived from our NFW fit, $c=5.4\pm0.2$ for a $M_{200}\sim 10^{15}
h_{50}^{-1} \msol$ cluster, to numerical simulations.  Teyssier
(\cite{teyssier}) derived c parameters in the range $4.9-9.5$ for 5
clusters in this very mass range.  The average $c$ parameter derived
by Eke \etal (\cite{eke}) for hot massive clusters is $c\sim 6$.  It
must be noted that a relatively large dispersion on this parameter is
expected from numerical simulations, with a $1\sigma$
$\Delta(\log{c})=0.18 (50\%)$ at a given mass (Bullock
\etal~\cite{bullock}).  In conclusion, we emphasize the excellent
agreement {\it in shape} between the mass profile derived for A1413
and the theoretical expectations, all the more remarkable in view of
the very small statistical errors on the profile.

\subsection{Normalisation of the $M$--$T$ relation}
\label{sec:mt}

We now examine the normalisation of the mass profile.  We will
classically define $M_{\delta}$, the mass within a given radius
$r_{\delta}$, inside which the mean mass density is $\delta$ times the
critical density, $\rho_{\rm c}(z)$ at the cluster redshift.  For
clusters obeying HE and self-similarity, the mass $M_{\delta}(T,z)$,
scales with the cluster temperature and redshift as:
\begin{equation}
      h(z) M_{\delta}(T,z) = M_{10}(\delta) T_{10}^{3/2}
\end{equation}
where $h(z)$ is the Hubble constant normalised to its local
value and $M_{10}(\delta)$ is the normalisation at density contrast
$\delta$ (here $T_{10}$ is arbitrarily expressed in unit of
$10~\kev$).  The above relation is remarkably well verified by
adiabatic numerical simulations, down to $\delta\sim 200$, for both
SCDM and $\Lambda{\rm CDM}$ cosmology,  with $M_{10}(\delta)$
independent of cosmology (e.g Mathiesen~\cite{math}).  The variation
of the normalisation $M_{10}(\delta)$ with $\delta$ reflects the
(universal) cluster internal structure and is the same for all
clusters for a given density contrast (although some scatter is
present in practice).

\begin{figure}
\begin{centering}
\includegraphics[scale=0.5,angle=0,keepaspectratio]{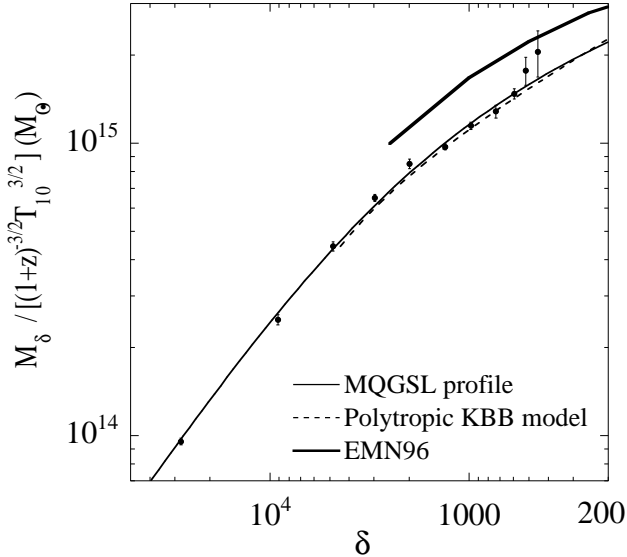}

\caption{{\footnotesize The scaled mass profile of A1413 (data
points), expressed as a function of density contrast, $\delta$,
compared with the simulations of Evrard \etal (1996) (thick line).
Filled circles: SCDM cosmology.  Open circles: $\Lambda$CDM cosmology.
The solid and dotted lines represent the scaled best-fit Moore \etal
(\cite{moore99}) profile and the scaled profile derived from the best
fit polytropic model, respectively (SCDM
cosmology).}}\label{fig:massdelta}
\end{centering}
\end{figure}

In Figure~\ref{fig:massdelta} we show the scaled mass profile of A1413
defined as:
 \begin{eqnarray} \widetilde{M}(\delta)& =& M(r)~h(z)
\left(\frac{\kTX}{10 \kev}\right)^{-3/2} \nonumber \\
   {\rm where} \nonumber\\
   \delta & =& \frac{3 M(r)}{4\pi r^{3} \rho_{\rm c}(z)}
    \label{eq:massdelta}
\end{eqnarray}
  with
   \begin{eqnarray}
    \rho_{\rm c}(z)& = & h^{2}(z) \frac{3 H_0^2}{ 8 \pi G}
    \label{eq:rhoc}
  \end{eqnarray}
and $h(z)=(1+z)^{3/2}$ for the SCDM cosmology considered here.  This
profile can be compared with the normalisations $M_{10}(\delta)$
derived from numerical simulations, allowing us to check the
normalisation of the $M_{\delta}$--$T$ relation at different density
contrasts.  The results of EMN96 are indicated with a thick line.  We
also plot the scaled profiles corresponding to our best-fit MQGSL
model (thin line), and to our best fit polytropic model (dashed line).
The density contrast is computed self-consistently for each profile
using Eq.~\ref{eq:massdelta}.

Both the data points and the MQGSL and polytropic model curves run
parallel to the EMN96 profile down to $\delta\sim 600$.  This simply
reflects the excellent agreement in shape of the A1413 profile with
numerical simulations, as outlined above.  However there is a very
significant offset in normalisation:
$\widetilde{M}(2500)=7.3\pm0.2\times 10^{14}~\msol$, compared with $
M_{10}(2500)=9.95 \times 10^{14}~\msol$ from EMN96.  In other words,
the predicted normalisation of the $M$--$T$ relation lies $\sim 36\%$
higher than the observed value, in excellent agreement with the {\it
Chandra\/} finding of Allen \etal (\cite{asf01}, $\sim 40\%$ at
$\delta=2500$).

Below $\delta=600$ the observed profile levels off, so that the data
points seem to converge towards the EMN96 predictions at small
$\delta$.  At $\delta=500$ $\widetilde{M}(500)=1.8^{+0.3}_{-0.2}
\times 10^{15}~\msol$, only $20 \%$ lower than the EMN96 value of $
M_{10}(500)=2.2 \times 10^{15}~\msol$ and actually consistent with it,
especially if we also take into account the dispersion observed in the
simulations ($15\%$).  If this effect is real, this would point to a
fundamental difference of {\it form\/} in the total mass profile at
large radii, i.e the real cluster dark matter density profile drops
less steeply than the canonical $r^{-3}$ law.  The observed
discrepancy in the normalisation of the $M$--$T$ relation at high
density contrast would thus be due to a flaw in the numerical
simulations for the dark matter component.

However, it is more likely that this level off of the observed
profile is an artifact due to incomplete virialisation.  If there is
residual kinetic energy due to infall, the HE equation applied to the
observed temperature profile would over-estimate the true mass.  Such
an effect is observed in the simulations of EMN96, although its
expected magnitude is somewhat smaller that we observe, $\sim 10\%$ at
$\delta=500$ on average (but a large scatter exists).  A further
indication that this is the correct explanation comes from a
comparison with the polytropic temperature model.  Let us assume that
the best fit MQGSL model indeed reflects the true total mass
distribution.  The fact that the mass profile derived from the best
fit polytropic model closely follows this profile down to $\delta=200$
(see Fig.~\ref{fig:massdelta}), indicates that this model is a true
representation of the thermodynamic state of the gas if it was in HE
up to there.  The drop in temperature in the last bin ($\delta> 550$),
as compared to this model, would thus be an direct indication of
incomplete thermalisation.  We also recall that there is a sudden drop
in the surface brightness profile at $\theta=7.8\arcmin$,
corresponding to a density contrast of 450 (computed with the MQGSL
model).  This further supports our interpretation: we might actually
be seeing the expected drop of the X-ray brightness beyond the edge of
the virialised (and hot) part of the cluster.  Finally, incomplete
equipartition between the electrons and the ions at the border of the
cluster (Chieze, Alimi \& Teyssier~\cite{chieze}) could also
contribute to the low (electronic) temperature observed.

The observed scaled mass profile depends on the cosmological
model via the function $h(z)$, used in the scaling, and the angular
distance, $d_{\rm A}$, used to convert angular radius to physical
radius.  On the other hand the theoretical normalisation
$M_{10}(\delta)$ appears to be insensitive to cosmology.  We thus
examined if a better agreement with the theoretical normalisation is
obtained for the currently most favoured $\Lambda$CDM model ($\Omo=0.3,
\Omega_\Lambda=0.7$).  From Eq.~\ref{eq:HE}, Eq.~\ref{eq:massdelta}
and Eq.~\ref{eq:rhoc}, the derived $\delta$ value scales as $(d_{\rm
A} h(z))^{-2}$ and $\widetilde{M}$ as $d_{\rm A} h(z)$, with
$h^{2}(z)=\Omo(1+z)^{3} +\Omega_\Lambda$.  For a $\Lambda$CDM model
(open circles in Fig.~\ref{fig:massdelta}), as compared to the SCDM
model (filled circles), the data points are moved down and left along
a line of slope 1/2 in the log-log plane, with $\delta$ multiplied by
1.137 and $\widetilde{M}$ multiplied by 0.938.  The translation is
modest and its slope is similar to the slope of the scaled mass
profile around $\delta=1000-500$, so the observed scaled mass profile
remains essentially unchanged and the agreement with the theoretical
curve is no better.

In summary the good agreement between the mass profile shape and the
numerical simulations, measured for the first time in the whole
virialised part of the cluster, suggest that the modelling of the Dark
Matter component is correct.  However, the offset in the normalisation
of the $M$--$T$ relation suggests that some physics is lacking in the
modeling of the gas.  Several groups have studied the effect of non
gravitational physics, like pre-heating or cooling, on the $M$--$T$
relation (e.g. Loewenstein~\cite{loewenstein}; Bialek
\etal~\cite{bialek}; Tozzi \& Norman~\cite{tozzi}; Babul
\etal~\cite{babul}; Thomas~\etal~\cite{thomas}; Voit
\etal~\cite{voit}).  A detailed comparison with numerical simulations,
which would require a statistically representative sample, is beyond
the scope of this paper.  We simply note that pure pre-heating seems
to have a small effect of the $M$-$T$ relation in the high temperature
range of A1413 (Loewenstein~\cite{loewenstein}; Tozzi \&
Norman~\cite{tozzi}; Babul \etal~\cite{babul}) and that models
including cooling (e.g. Thomas~\etal~\cite{thomas}) seem to be more
successful.  We also emphasize that some care must be exercised when
comparing theoretical predictions with these X--ray observations.  The
magnitude of the effect is not large as compared to the typical
difference of $\sim 50\%$ (Henry~\cite{henry}) in the normalisation
derived by various groups using purely adiabatic simulations.  The
dispersion of the relation, observed for both simulated ($\sim 20\%$)
and real clusters, require the use of statistically well controlled
samples.  Finally, a further ambiguity lies in the definition of the
temperature.  The temperature profile is not exactly isothermal
(although our data suggest that the departure is small).  Ideally we
should compare the data from a given instrument using the spectral
temperature estimated from simulation, after full modelling of the
plasma emission folded with the instrument response.  The study of
Mathiesen \& Evrard (\cite{me01}) indicates that the spectral
temperature could be an underestimate of the mass-weighted temperature,
by about $20\%$.  Note that this effect would worsen the discrepancy
observed above.  All these systematic effects have to be well
controlled, if we want to confirm the departure from the self-similar
scaling and identify the physical process responsible for it.

\begin{figure}
\begin{centering}
\includegraphics[scale=0.5,angle=0,keepaspectratio]{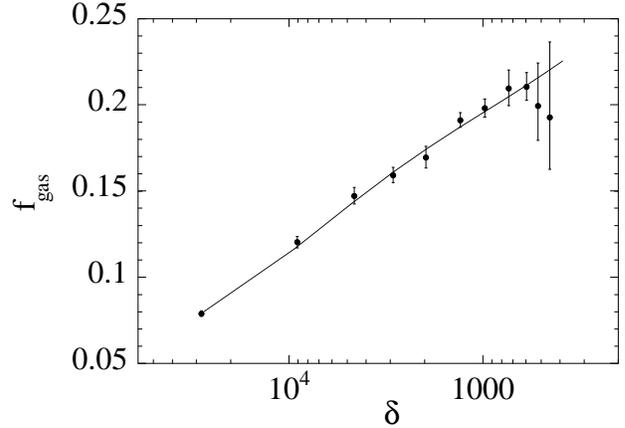}

\caption{{\footnotesize The gas mass fraction of A1413 as a function
of overdensity $\delta$.  Data points: gas mass fraction obtained
using the total mass derived from the Monte-Carlo method.  Line: same
using the best fit MQGSL mass profile. }}\label{fig:figfgas}
\end{centering}
\end{figure}

\subsection{Gas distribution}
\label{sec:dissgasdens}

  From our determination of the central gas density we have calculated
a cooling radius of $r_{\rm cool} = 0'.6$.
Figure~\ref{fig:FigProffit} shows that the gas density has begun to
rise above the \betamodel\ fit at about twice this radius. This
suggests that
the main driver of the central peak in the gas distribution is the
cusp in the dark matter profile, not the CF. It is likely that the
cusp in the dark matter profile acts as a focus for the gas and the
CF. This has implications for CF mass deposition rates $\dot{M}$
deduced from excesses to the \betamodel, in that any application to
this cluster would result in a gross overestimate of $\dot{M}$.

The ratio of the X-ray gas mass to the total gravitating mass is shown
as a function of density contrast $\delta$ in
Figure~\ref{fig:figfgas}.  The $\fg$ rises slowly with increasing
radius, up $\delta \sim 1000$.  After this point, the $\fg$ either
stabilises within the error if we use the total mass data points, or
continues increasing beyond if we used the best fit MQGSL total mass
profile.  A conservative estimate of the gas mass fraction at
$\delta\sim500$ is $\fg=0.2\pm0.02$.  This shows that the hot ICM is
more extended than the dark matter distribution, as has been found
in previous studies (e.g., David, Jones \&
Forman~\cite{djf95}), and is also seen in numerical simulations, and
expected from purely dynamical reasons (Chieze, Teyssier \&
Alimi~\cite{chieze97}).

Assuming that the gas density profile follows the dark matter profile
at large radii (within a factor of 2 between $r_{200}/2$ and
$2r_{200}$) and a polytropic equation of state, Komatsu \& Seljak
(\cite{komatsu}) derived an analytical solution of the gas in HE in an
NFW profile.  They obtained $\gamma=1.15 +0.01(c-0.5)$, or
$\gamma=1.14$ for the concentration parameter $c=5.4$ derived for
A1413, and a X-ray outer slope of $\beta\sim 0.65$ for a typical 6.5
keV cluster (their Fig.  14).  The first assumption is roughly
verified for our best fit profiles and not surprisingly, our derived
values are in fair agreement with their model, although we obtain a
slightly but significantly lower value of $\gamma=1.07\pm 0.01$ and
a slightly higher $\beta$ value.

The value at $\delta=600$, corresponding roughly to the virialised
part of the cluster, $\fg \simeq 0.2$, can be used to calculate the
total mass
density in the Universe, following the arguments that assume that the
properties of clusters constitute a fair sample of those of the
Universe as a whole (e.g., White \etal \cite{whi93}).  Assuming that
the
luminous baryonic mass in galaxies in A1413 is approximately one-fifth
of the X-ray ICM mass (e.g., White \etal \cite{whi93}), and
neglecting other possible sources of baryonic dark matter, $\Omega_m =
(\Omega_b/1.2 \fg)$, where $\Omega_b$ is the mean baryon density.  For
$\Omega_b h_{50}^2 = 0.0820$ (O'Meara \etal~\cite{ome}, $h_{50}$ is
the Hubble
constant in units of 50 km s$^{-1}$ Mpc$^{-1}$), we obtain $\Omega_m =
0.34 h_{50}^{-0.5}$.



\section{Conclusions}

The main conclusions of this work may be summarised as follows:

\begin{enumerate}

\item We have reported \xmm\ observations of A1413, a relaxed galaxy
cluster at z=0.143.

\item In a 2D $\beta$ model analysis, we
detect substructure to the south which does not appear to be
interacting with the main cluster.

\item Excluding the data from this region and all obvious point
sources, we have measured the gas density and temperature profiles up
to $r_{500}$ (corresponding to a density contrast $\delta \sim 500$,
with respect to the critical density at the redshift of the cluster).
With the assumptions of HE and spherical symmetry, we have calculated
the mass profile out to the same distance.

\item The gas density profile is well described with a $\beta$-model
beyond $\sim 250 h_{50}^{-1}~{\rm kpc}$.  We further parameterise the
inner regions with a modified version of the $\beta$-model (the KBB,
Eq.~\ref{eq:kbb}), which allows a more centrally peaked gas distribution.

\item The temperature profile (excluding the inner point) is well
described by a polytropic model with $\gamma = 1.07\pm 0.01$.  The
decline is modest: a decrease of $\sim 20\%$ between $0.1 r_{200}$ and
$0.5 r_{200}$.

\item The mass profile, derived from the HE equation, is determined
with an accuracy of about $\pm5\%$ up to $r_{600}$ and $\sim \pm 18\%$
at $r_{500}$.  It can be remarkably well described by a Moore
\etal~(\cite{moore99}) profile with a scale radius $\rs = 845 \pm 43$
kpc and concentration parameter $c = 2.6 \pm 0.1$.  An NFW profile
also gives an acceptable fit but describes less well the central
regions.  The $c$ values we find are in good agreement with those
expected from numerical simulations for a cluster of this mass.  The
Dark Matter modelling in these simulations is thus strongly supported
by the excellent agreement between observed and simulated profiles.

\item Beyond $r_{600}$, the observed temperature and derived mass
profiles begin to depart systematically from, respectively, the
polytropic description and the Moore \etal (\cite{moore99}) profile.
There is also a sudden drop of the surface brightness profile at
$r_{450}$.  This suggests that the gas in these regions may not be in
HE, and we may thus be seeing the outer edge of the virialized parts
of this cluster.

\item The offset in the normalisation of the $M_{\delta}-T$
relation, with respect to
the simulations of Evrard et al. (1996) is now confirmed to be $\sim
40\%$ across the entire radial range up to $r_{500}$ (i.e., in the
virialised part of the
   cluster).

\item The gas distribution is peaked primarily as a result of the
cusp in the dark matter profile. The gas mass fraction increases with
increasing radius, to reach $\sim 0.2$ at $r_{500}$.
\end{enumerate}

We are now in a position directly to confront simulations with
observations. The results are encouraging (the obvious validity of
the modelling of the Dark Matter distribution at large scale) but many
questions remain.  How peaked are dark matter profiles?  What is the
relationship between between central dark matter cusps and CFs?  Why
are some studies finding unrealistic values of the concentration
parameter?  What is the source of the discrepancy in the M-T relation?

The statistical errors on the observed quantities are now small enough
so that we can determine in detail the intrinsic dispersion in cluster
properties and systematic discrepancies with the classical
self-similar
model.  To answer the above and other questions, a statistical sample
of cluster properties would be of great help, preferably using {\it
Chandra\/} to probe the central regions and \xmm\ to determine
properties at great distances from the cluster centre.  Confrontation
with numerical simulation is essential.  The full range of
observations, correlations between cluster properties, and detailed
internal gas structure should be derived taking into account that they
are viewed through a given instrument, so that we are able truly to
compare like with like.


\begin{acknowledgements}

We thank the referee, J. Irwin, for insightful comments and
suggestions which have improved the paper.  We thank T. Ponman for
providing the script used in the deprojection of the temperature
profile, and D. Neumann for providing the Monte-Carlo code to derive
the mass profile.  We thank R. Teyssier and A. Refregier for useful
discussions, and S. De Grandi for providing the {\it BeppoSAX\/}
results shown in Fig.~\ref{fig:tempcomp}.  We thank A.Evrard,
M.Markevitch and T.Ponman for their comments on the manuscript.  The
present work is based on observations obtained with \xmm, an ESA
science mission with instruments and contributions directly funded by
ESA Member States and the USA (NASA).

\end{acknowledgements}


\appendix
\section{Cleaning of data for soft proton solar flares}

Various methods have been used to remove background flares from \xmm\
and
{\it Chandra\/} data sets.  For \xmm\ observations, perhaps the most
simple is direct visual inspection of the binned high-energy [10-12]
keV light curve over the whole field of view, the adoption of a
threshold level, and the exclusion of any intervals above the selected
threshold (see e.g., Arnaud et al.  2001).  For {\it Chandra\/} data,
an $n\sigma$ clipping is used, where the average count rate in the [3
- 6] keV band over the field of view is calculated, thresholds are set
depending on this value, and the light curve cleaned for any intervals
where the thresholds are not met.

It is known that the \xmm\ quiescent background level is variable by
$\sim \pm 10\%$, and so it is not possible to set a rigid threshold
level for flare rejection because of the risk of losing good data.
The threshold level should ideally be dependent on the quiescent rate
of the observation in question.

\begin{figure}
\begin{centering}
\includegraphics[scale=0.5,angle=0,keepaspectratio]{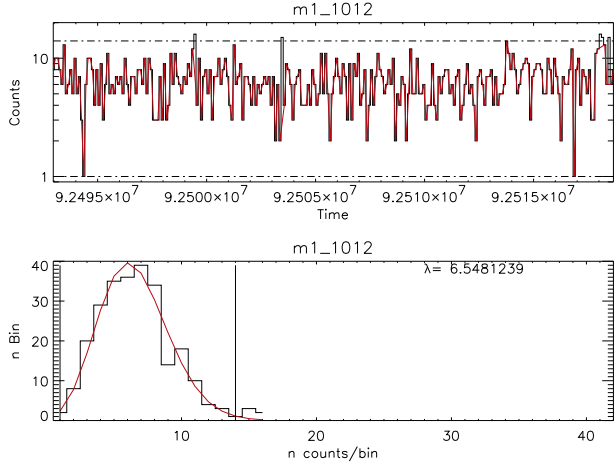}
\caption{{\footnotesize Top: The \mos1 [10-12 keV] light curve of
A1413 before (black) and after (red) cleaning for flares. Dashed
lines show the $\pm 3 \sigma$ thresholds. Bottom: The Poisson fit to
the histogram of the light curve, from which the thresholds are
calculated. The upper threshold is 14 cts/104s.}}\label{fig:lcrv}
\end{centering}
\end{figure}

\begin{figure}
\begin{centering}
\includegraphics[scale=0.5,angle=0,keepaspectratio]{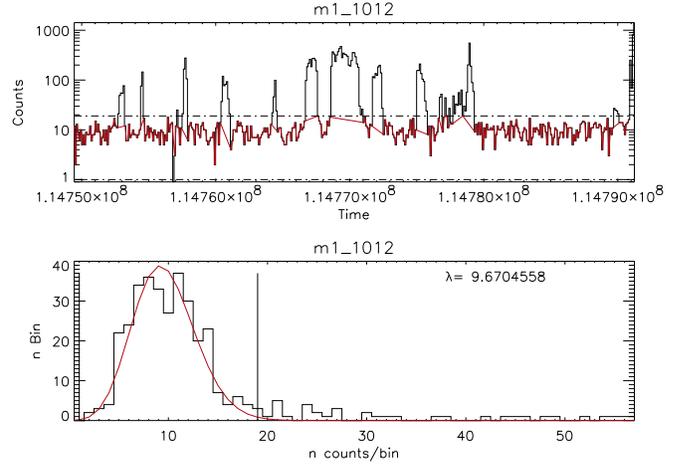}
\caption{{\footnotesize Top: The \mos1 [10-12 keV] light curve of
MKW9 before (black) and after (red) cleaning for flares. Dashed lines
show the $\pm 3 \sigma$ thresholds. Note the y-axis is in log units.
Bottom: The Poisson fit to the histogram of the light curve, from
which the thresholds are calculated. The x-axis has been zoomed to
emphasise the fit. The upper threshold is 19 cts/104s.
}}\label{fig:lcrv2}
\end{centering}
\end{figure}

For the A1413 observation, we extracted the \mos [10-12] keV light
curves in the field of view in 104s bins (chosen as this is an
integral multiple of the frame readout time). Similar light curves
were made for the pn, but in the [12-14] keV band. We then made a
histogram of each light curve and fitted this histogram with a
Poisson distribution

\begin{equation}
y = \frac{\lambda^x e^{-\lambda}}{x!}
\end{equation}

\noindent where the mean of the distribution, $\lambda$, is the free
parameter of the fit. Following Poisson statistics, the error on the
mean, $\sigma = \sqrt{\lambda}$. We found that the Poisson
distribution was, without exception, a better fit than a Gaussian
distribution. We then defined thresholds at the $\pm 3\sigma$ level,
and rejected any time intervals outside these thresholds. We show the
Poisson fit and the original and cleaned light curves for A1413,
which is a quiet observation, and for comparison, the observation of
MKW9 (Neumann et al., in preparation), which shows several solar
flares, in Figures~\ref{fig:lcrv} and Figures~\ref{fig:lcrv2}.

This method is effective at finding the quiescent periods,
even for data strongly affected by flares. As such it is not prone to
the overestimation of the mean rate, the main problem with the
$n \sigma$ clipping method. 


\section{Table}

\begin{sidewaystable}[h]
\centering
\caption{{\small Radial temperature profile results. The spectrum of
each annulus has been binned to $3\sigma$ significance, except the
final annulus, which has been binned to $2\sigma$. They are then
fitted with a MEKAL model assuming a fixed absorption of $N_H = 2.19
\times 10^{20}$ cm$^{-2}$. Errors are given at $90\%$ confidence for
one interesting parameter.
Note that some abundance measurements have been frozen at the global
value found for the cluster.}}
\begin{tabular}{l l l l l l l l l l l l l}
\hline

\multicolumn{1}{l}{ Annulus } & \multicolumn{3}{l}{ {\mos1} } &
\multicolumn{3}{l}{ {\mos2} } & \multicolumn{3}{l}{ pn } &
\multicolumn{3}{l}{ {\sc MOS}+pn } \\

\multicolumn{1}{l}{  ( $\arcmin$ ) } & \multicolumn{1}{l}{ $kT$
(keV)} & \multicolumn{1}{l}{ $Z$ ($Z_{\odot}$) } &
\multicolumn{1}{l}{ $\chi^2$ / $\nu$} & \multicolumn{1}{l}{ $kT$
(keV)} & \multicolumn{1}{l}{ $Z$ ($Z_{\odot}$) } &
\multicolumn{1}{l}{ $\chi^2$ / $\nu$} & \multicolumn{1}{l}{ $kT$
(keV)} & \multicolumn{1}{l}{ $Z$ ($Z_{\odot}$) } &
\multicolumn{1}{l}{ $\chi^2$ / $\nu$} & \multicolumn{1}{l}{ $kT$
(keV)} & \multicolumn{1}{l}{ $Z$ ($Z_{\odot}$) } &
\multicolumn{1}{l}{ $\chi^2$ / $\nu$}\\

\hline

    0.00 - 0.50
& $ 7.25^{+ 0.58}_{- 0.50} $ & $ 0.36^{+ 0.13}_{- 0.12} $ & 216.2/276
& $ 6.13^{+ 0.49}_{- 0.43} $ & $ 0.35^{+ 0.11}_{- 0.10} $ & 274.3/266
& $ 7.05^{+ 0.86}_{- 0.72} $ & $ 0.31^{+ 0.14}_{- 0.13} $ & 259.9/301
& $ 6.75^{+ 0.31}_{- 0.31} $ & $ 0.34^{+ 0.07}_{- 0.07} $ & 758.5/848
\\

    0.50 - 0.69
& $ 7.72^{+ 1.02}_{- 0.71} $ & $ 0.37^{+ 0.19}_{- 0.17} $ & 205.8/208
& $ 6.67^{+ 0.71}_{- 0.70} $ & $ 0.36^{+ 0.17}_{- 0.16} $ & 194.8/205
& $ 6.96^{+ 1.22}_{- 1.01} $ & $ 0.27^{+ 0.20}_{- 0.19} $ & 141.1/172
& $ 7.14^{+ 0.48}_{- 0.45} $ & $ 0.33^{+ 0.10}_{- 0.10} $ & 550.0/590
\\

    0.69 - 0.97
& $ 7.03^{+ 0.70}_{- 0.61} $ & $ 0.37^{+ 0.15}_{- 0.15} $ & 238.8/227
& $ 6.65^{+ 0.63}_{- 0.62} $ & $ 0.25^{+ 0.13}_{- 0.13} $ & 236.3/228
& $ 7.66^{+ 1.36}_{- 1.03} $ & $ 0.33^{+ 0.19}_{- 0.17} $ & 174.2/231
& $ 6.97^{+ 0.40}_{- 0.40} $ & $ 0.31^{+ 0.09}_{- 0.08} $ & 654.1/691
\\

    0.97 - 1.35
& $ 7.31^{+ 0.72}_{- 0.58} $ & $ 0.43^{+ 0.16}_{- 0.15} $ & 257.3/238
& $ 6.72^{+ 0.59}_{- 0.58} $ & $ 0.27^{+ 0.12}_{- 0.12} $ & 281.8/249
& $ 7.02^{+ 0.98}_{- 0.80} $ & $ 0.32^{+ 0.16}_{- 0.15} $ & 237.0/254
& $ 7.00^{+ 0.37}_{- 0.36} $ & $ 0.33^{+ 0.08}_{- 0.08} $ & 779.4/746
\\

    1.35 - 1.89
& $ 6.35^{+ 0.55}_{- 0.51} $ & $ 0.40^{+ 0.14}_{- 0.13} $ & 209.7/243
& $ 6.49^{+ 0.57}_{- 0.56} $ & $ 0.34^{+ 0.13}_{- 0.13} $ & 228.0/242
& $ 7.10^{+ 1.02}_{- 0.79} $ & $ 0.33^{+ 0.16}_{- 0.15} $ & 228.5/289
& $ 6.55^{+ 0.35}_{- 0.35} $ & $ 0.36^{+ 0.08}_{- 0.08} $ & 673.8/779
\\

    1.89 - 2.64
& $ 6.57^{+ 0.65}_{- 0.63} $ & $ 0.18^{+ 0.12}_{- 0.12} $ & 187.7/222
& $ 6.41^{+ 0.63}_{- 0.57} $ & $ 0.24^{+ 0.13}_{- 0.12} $ & 201.7/224
& $ 7.44^{+ 1.42}_{- 0.94} $ & $ 0.23^{+ 0.16}_{- 0.15} $ & 249.9/241
& $ 6.67^{+ 0.41}_{- 0.40} $ & $ 0.22^{+ 0.08}_{- 0.07} $ & 642.2/692
\\

    2.64 - 3.68
& $ 6.87^{+ 0.85}_{- 0.78} $ & $ 0.27 $ frozen& 201.5/190
& $ 5.64^{+ 0.67}_{- 0.52} $ & $ 0.27$ frozen & 217.4/186
& $ 6.33^{+ 1.19}_{- 1.06} $ & $ 0.27$ frozen & 121.2/161
& $ 6.22^{+ 0.51}_{- 0.46} $ & $ 0.26^{+ 0.10}_{- 0.10} $ & 543.4/539
\\

    3.68 - 5.13
& $ 6.27^{+ 1.16}_{- 0.91} $ & $ 0.27$ frozen & 128.0/137
& $ 5.83^{+ 1.16}_{- 0.77} $ & $ 0.27$ frozen & 144.2/133
& $ 5.72^{+ 2.19}_{- 1.20} $ & $ 0.27$ frozen & 100.9/90
& $ 6.01^{+ 0.78}_{- 0.61} $ & $ 0.18^{+ 0.17}_{- 0.16} $ & 371.6/362
\\

    5.13 - 7.16
& $ 5.45^{+ 2.06}_{- 1.27} $ & $ 0.27$ frozen & 42.7/57
& $ 5.59^{+ 2.93}_{- 1.48} $ & $ 0.27$ frozen & 61.2/47
& $ 5.35^{+ 6.81}_{- 2.15} $ & $ 0.27$ frozen & 22.5/29
& $ 5.67^{+ 1.56}_{- 1.16} $ & $ 0.50^{+1.62}_{-0.50}$ & 133.3/135 \\

    7.16 - 9.00


& $ 3.50^{+ 3.71}_{- 1.25} $ & $ 0.27$ frozen & 15.9/18
& - & - &  -
& $ 2.86^{+ 97.14}_{- 21.6} $ & $ 0.27$ frozen & 1.8/4
& $ 4.13^{+ 4.61}_{- 2.08} $ & $ 0.00^{+1.97}_{-0.00}$ & 37.0/36 \\

\hline
\end{tabular}
\label{tab:ktres}
\end{sidewaystable}

\begin{table}[hb]
\centering
\center
\caption{{\small Deprojected temperature profile results, with
$3\sigma$ errors. \mos and pn spectra are fitted as described in
Sect.~\ref{sec:deproj}.}}
\begin{tabular}{ l l l l }
\hline

\multicolumn{1}{l}{ Annulus } & \multicolumn{1}{l}{ kT } &
\multicolumn{1}{l}{ Abundance } \\
\multicolumn{1}{l}{ ( ' )  } & \multicolumn{1}{l}{ (keV) } &
\multicolumn{1}{l}{ $Z/Z_{odot}$ } \\

\hline

0.00 - 0.50 & $6.40^{+0.47}_{-0.47}$ & $0.34^{+0.06}_{-0.06}$ \\
0.50 - 0.69 & $7.83^{+1.97}_{-1.47}$ & $0.38^{+0.21}_{-0.20}$ \\
0.69 - 0.97 & $6.69^{+1.20}_{-0.95}$ & $0.29^{+0.15}_{-0.13}$ \\
0.97 - 1.35 & $7.70^{+1.29}_{-1.00}$ & $0.24^{+0.12}_{-0.11}$ \\
1.35 - 1.89 & $6.39^{+0.67}_{-0.59}$ & $0.52^{+0.10}_{-0.10}$ \\
1.89 - 2.64 & $6.89^{+0.81}_{-0.68}$ & $0.17^{+0.08}_{-0.07}$ \\
2.64 - 3.68 & $6.35^{+0.84}_{-0.71}$ & $0.34^{+0.10}_{-0.10}$ \\
3.68 - 5.13 & $6.09^{+1.20}_{-0.87}$ & $0.11^{+0.24}_{-0.11}$ \\
5.13 - 7.16 & $6.89^{+4.35}_{-2.06}$ & 0.3 frozen \\
7.16 - 9.00 & $4.13^{+4.61}_{-2.08}$ & 0.3 frozen\\

\hline
\end{tabular}
\label{tab:deproj}
\end{table}


\clearpage

\begin{thebibliography}{}


\bibitem[2001a]{asf01}
          Allen S.W., Schmidt, R.W., Fabian, A.C., 2001, MNRAS, 328,
L37

\bibitem[2001b]{asf01b}
          Allen S.W., Schmidt, R.W., Fabian, A.C., 2001, MNRAS, in
press,  astro-ph/0111368

\bibitem[2002]{abg02}
          Arabadjis, J.S., Bautz, M.W., Garmire, G.P., 2002, ApJ, 572, 66

   \bibitem[2001a]{arnaud01}
       Arnaud, M., Aghanim, N., Gastaud, R. et al.  2001a, A\&A, 365, L67

   \bibitem[2001b]{maetal01}
  Arnaud, M., Neumann, D., Aghanim, N., et al. 2001b, A\&A, 365, L80

\bibitem[2002]{manadmn02}
         Arnaud, M., Aghanim, N., Neumann, D., 2002, A\&A, 389, 1


\bibitem[2002]{maetal02} Arnaud, M., Majerowicz, S., Lumb, D., \etal,
2002, A\&A, in press, astro-ph/0204306

\bibitem[2002]{babul} Babul, A., Balogh, M.L, Lewis, G.F., Poole, G.B,
2002, MNRAS, 330, 329

\bibitem[1999]{bialek} Bialek, J.J., Evrard, A.E., Mohr, J.J. 2001,
ApJ, 555, 597

\bibitem[2001]{bullock} Bullock, J.S., Kollat, T.S., Sigad, Y. \etal
2001, MNRAS, 321, 559

\bibitem[2001]{buote01}
      Buote, D., 2001, Proc. Merging Processes in Clusters of Galaxies,
eds. L.Ferretti, I.M. Gioia \& G. Giovannini (Kluwer, Dodrecht)

\bibitem[1997]{chieze97} Chieze, J.P.,  Teyssier, R., Alimi, J.M.,
1997, ApJ,
484, 40

\bibitem[1998]{chieze} Chieze, J.P., Alimi, J.M., Teyssier, R., 1998,
ApJ,
495, 630

\bibitem[1997]{cnt97}
         Cirimele, G., Nesci, R., Trevese, D., 1997, ApJ, 475, 11

\bibitem[1995]{djf95}
         David, L.P., Jones, C., Forman, W., 1995, ApJ, 445, 578

         \bibitem[2001]{david01}
         David, L.P., Nulsen, P.E.J., McMamara, B.R., Forman, W.,
Jones, C., Ponman, T., Robertson, B., Wise, M., 2001, ApJ, 557, 546

\bibitem[2001]{dm01}
         De Grandi, S., Molendi, S., 2001, ApJ, 551, 153

\bibitem[2002]{demol02}
         De Grandi, S., Molendi, S., 2002, ApJ, 567, 163 (DM02)

\bibitem[1990]{dickey} Dickey, J.M., Lockman, F.J. 1990, ARA\&A, 28,
215

\bibitem[2002]{durret} Durret, F., Slezak E., Lieu, R., Dos Santos,
S.,
Massimiliano Bonamente, M., 2002, A\&A, in press, astro-ph/0204345

\bibitem[1998]{eke} Eke, V.R., Navarro, J.F., Frenk, C.S. 1998, ApJ,
503, 569

\bibitem[1995]{elbaz}Elbaz, D., Arnaud, M., B\"ohringer H. 1995, A\&A,
293, 337

\bibitem[1996]{emn96}
         Evrard, A.E., Metzler, C.A., Navarro, J.F., 1996, ApJ, 469,
494 (EMN96)

\bibitem[2001]{frb01}
         Finoguenov, A., Reiprich, T.H., B\"{o}hringer, H., 2001, A\&A,
368, 749

\bibitem[2001]{ghizzardi} Ghizzardi, S. 2001, EPIC-MCT-TN-011
(XMM-SOC-CAL-TN-0022)

\bibitem[2002]{griffiths02a} Griffiths, R.G., Briel, U., Dadida, M.,
\etal 2002, in proceedings of the "New Vision of the X-ray Universe
in the
XMM-Newton and Chandra Era" conference, to appear

\bibitem[2002]{griffiths02b} Griffiths, G., Saxton, R. 2002, in
preparation

\bibitem[2000]{henry} Henry, J.P., 2000, ApJ, 534, 565

\bibitem[1999]{hms99}
         Horner, D.J., Mushotzky, R.F., Scharf, C.A., 1999, ApJ, 520,
78

\bibitem[2002]{ikebe02}
         Ikebe, Y., Reiprich, T.H., B\"{o}hringer, H., Tanaka, Y.,
Kitayama, T., 2002, A\&A, 383, 773

\bibitem[2000]{irbreg01}
         Irwin, J.A., Bregman, J., 2000, ApJ, 538, 543 (IB00)

\bibitem[2001]{ib01}
         Irwin, J.A., Bregman, J., 2001, ApJ, 546, 150

\bibitem[2001]{komatsu} Komatsu, E., Seljak, U., 2001, MNRAS, 327,
1553

\bibitem[2000]{loewenstein} Lowenstein, M., 2000, ApJ, 532, 17

\bibitem[2002]{lumb}
Lumb, D. 2002, XMM-SOC-CAL-TN-0016

\bibitem[2002]{majerow} Majerowicz, S., Neumann, D.M., Reiprich, T.H.,
2002, A\&A, submitted, astro-ph/0202347

\bibitem[1998]{mark98}
         Markevitch, M., Forman, W., Sazazin, C., Vikhlinin, A., 1998,
ApJ, 503, 77 (MFSV98)

\bibitem[1999]{mark99} Markevitch M., Sarazin C., Vikhlinin A., 1999,
ApJ,
521, 526

\bibitem[2002]{mark02} Markevitch, M., 2002, Technical Note,
astro-ph/0205333

\bibitem[2001]{me01} Mathiesen, B.F., Evrard, A.E. 2001, ApJ, 546, 100

\bibitem[2001]{math} Mathiesen, B.F., 2001,MNRAS, 326, L1


\bibitem[2001]{matsu01}
         Matsumoto, H., Tsuru, T., Fukazawa, Y., Hattori, M., Davis,
D.,
2001, PASJ, 52, 153

\bibitem[1999]{moore99}
         Moore, B., Quinn, T., Governato, F., Stadel, J., Lake, G.,
1999, MNRAS, 310, 1147 (MQGSL)

\bibitem[1997]{nfw97}
        Navarro, J.F., Frenk, C.S., White, S.D.M., 1997, ApJ, 490, 493
(NFW)

\bibitem[2000]{nmf00}
        Nevalainen, J., Markevitch, M., Forman, W., 2000, ApJ, 536, 73

\bibitem[1995]{neubo95}
         Neumann, D., B\"{o}hringer, H., 1995, A\&A, 301, 865

\bibitem[1997]{neuboh97}
         Neumann, D., B\"{o}hringer, H., 1997, MNRAS, 289, 123

\bibitem[1999]{dmnma99}
         Neumann, D., Arnaud, M., 1999, A\&A, 348, 711

\bibitem[2001]{dmnetal01}
         Neumann, D., Arnaud, M., Gastaud, R., et al. 2001, A\&A, 365,
L74

\bibitem[2001]{ome} O'Meara, J.M., Tytler, D., Kirkman, D. \etal,
2001, ApJ, 552, 718

\bibitem[1994]{oh94}
       Oegerle, W.R., Hill, J.M., 1994, AJ, 107, 857

\bibitem[1995]{ohf95}
      Oegerle, W.R., Hill, J.M., Fitchett, M.J., 1995, AJ, 110, 32

\bibitem[1998]{p98}
         Peres, C.B., Fabian, A.C., Edge, A.C., Allen, S.W., Johnstone,
R.M., White, D.A., MNRAS, 298, 416

\bibitem[1980]{perr80}
         Perrenod, S.C., 1980, ApJ, 236, 373

\bibitem[2001]{pratt}
         Pratt, G.W., Arnaud, M., Aghanim, N., 2001, Proc. XXXVI
Rencontres de Moriond:  Galaxy Clusters and the High-Redshift
Universe, eds. D.M. Neumann, F. Durret and J. Tr\^{a}n Thanh Van:
{astro-ph/0105431}.

\bibitem[1986]{sarazin} Sarazin, C.  1986, Reviews of Modern Physics,
58, 1

\bibitem[2002]{saxton} Saxton, R.  2002, XMM-SOC-CAL-TN-0023

\bibitem[1996]{schindler96}
         Schindler, S., 1996, A\& A, 305, 756

\bibitem[1997]{snow97}
         Snowden, S., Egger, R., Freyberg, M.J., et al. 1997, ApJ, 485,
125

\bibitem[1998]{suto98}
         Suto, Y., Saski, S., Makino, N., 1998, ApJ, 509, 544


\bibitem[2002]{teyssier} Teyssier, R., 2002, A\&A, 337, 364

\bibitem[2002]{thomas} Thomas, P.E., Muanwong, O., Kay, S.T., Liddle,
A.R., 2002, MNRAS, 330, L48

\bibitem[2001]{tozzi} Tozzi P., Norman C., 2001, ApJ 546, 63

\bibitem[1999]{vikh99}
         Vikhlinin, A., Forman, W., Jones, C., 1999, ApJ, 525, 47

\bibitem[2002]{voit}
         Voit, G.M., Bryan, G.L., Balogh, M.L., Bower, R.G., 2002, ApJ,
in
         press, astro-ph/0205240

\bibitem[1993]{whi93}
         White, S.D., Navarro, J.F., Evrard, A.E., Frenk, C.S., 1993,
Nature, 366, 429

\bibitem[2000]{whi00}
         White, D.A., 2000, MNRAS, 312, 663


\end{thebibliography}

\end{document}